\documentclass[journal]{IEEEtran}

\usepackage[numbers]{natbib}
\usepackage{graphicx}
\usepackage{mathtools}
\graphicspath{{images/}}
\usepackage[style=base]{caption}
\usepackage{subcaption}
\usepackage{amsmath}
\usepackage{multirow}
\usepackage{float}
\graphicspath{{images/}}
\usepackage{adjustbox}
\usepackage[normalem]{ulem}

\begin{document}

\title{File Fragment Classification using Light-Weight Convolutional Neural Networks}

\author{Mustafa Ghaleb, Kunwar Saaim, Muhamad Felemban~\IEEEmembership{Member,~IEEE,}, Saleh Al-Saleh, Ahmad Al-Mulhem 
\thanks{M. Ghaleb, M. Felemban, S. Al-Saleh, and A. Al-Mulehm are in the Department
of Computer Engineering and the Interdisciplainry Research Center for Intelligent Secure Systems at KFUPM, Dhahran, Saudi Arabia}
\thanks{K. Saaim is with the Department of Computing Science, University of Alberta}}

\maketitle

\begin{abstract}
In digital forensics, file fragment classification is an important step toward completing file carving process. There exist several techniques to identify the type of file fragments without relying on meta-data, such as using features like header/footer and N-gram to identify the fragment type. Recently, convolutional neural network (CNN) models have been used to build classification models to achieve this task. However, the number of parameters in CNNs tends to grow exponentially as the number of layers increases. This results in a dramatic increase in training and inference time. In this paper, we propose light-weight file fragment classification models based on depthwise separable CNNs. The evaluation results show that our proposed models provide faster inference time with comparable accuracy as compared to the state-of-art CNN based models. In particular, our models were able to achieve an accuracy of 79\% on the FFT-75 dataset with nearly 100K parameters and 164M FLOPs, which is 4x smaller and 6x faster than the state-of-the-art classifier in the literature.\end{abstract}


%
\IEEEpeerreviewmaketitle

\section{Introduction}

Digital forensics is the science of collecting digital evidence to investigate cybercrimes and cyberattacks. The process of digital forensics includes preservation, identification, and extraction of files and data from damaged or compromised machines. There are several digital forensics techniques and tools that facilitate this process \cite{rafique2013exploring,marziale2007massive}. Often, attackers attempt to wipe out any evidence that incriminates them, for example, by formatting the hard disk  \cite{bennett2012challenges}. In such scenarios, traditional recovery methods based on the file system meta-data are deemed ineffective. 

To overcome this challenge, digital investigators use file carving to reconstruct files based on their contents \cite{garfinkel2007carving, lei2011forensic, lin2018file}. File carving process recovers damaged data files from blocks of raw binary data without using any meta-data in the file system. There are several file carving techniques to reconstruct files fully or partially using a variety of techniques including header/footer matching \cite{richard2005scalpel}, probabilistic measures \cite{alghafli2019techniques}, and n-gram analysis \cite{wang2018sparse}. 

Without meta-data, it is challenging to identify the type of a carved file. The problem is even more challenging when file carving is carried out on fragmented files. File fragmentation usually occurs when there is not enough contiguous space to write a large file on the hard disk. Two tasks can be recognized in fragmented file carving: selecting a candidate sequence of file blocks and classifying the type of the fragment \cite{lin2018file}. The latter is called file fragment classification. 


There exist several approaches for file fragment classification in the literature \cite{wang_su_song_2018,mittal2020fifty,beebe2013sceadan,wang2018sparse,richard2005scalpel,chen_hui_2018,Q_Li_SVM,fitzgerald_mathews_morris_zhulyn_2012,ahmed_lhee_shin_hong_2011,Amirani2013FeaturebasedTI,Zheng2015AFC,vulinovic2019neural,Hiester2018FileFC}. For example, \textit{Sceadan} tool determines the type of bulk data based file content \cite{beebe2013sceadan}. Recently, deep learning has been used to identify the type of file fragment. For example, Gray-scale \cite{chen_hui_2018} and FiFTy \cite{mittal2020fifty} tools illustrate how CNNs can be efficiently used for file fragment classification. Although CNN is the most used deep learning model for classification, we argue that existing models can be further improved in terms of performance and accuracy by modifying the architecture of the neural network. We particularly focus on light-weight CNNs to be able to carry out file fragment classification on resource-limited devices.  

A major drawback of CNNs is the exponential increase in the number of parameters as the number of layers increases, i.e., the model depth increases. Such an increase in the number of parameters increases the complexity of the model and, accordingly, increases the training and inference time. On the other hand, deep CNN models provide better accuracy. Therefore, it is important to find a balance between inference speed and accuracy by optimizing the CNN structure, i.e., number of layers, CNN filters size, and number of channels. To overcome this challenge, researchers proposed a computationally cheaper model called depthwise separable convolution \cite{mobilenet,inception,tan2019efficientnet}, in which the number of parameters in the model is reduced dramatically. 

In our previous paper \cite{saaim2022light}, we proposed a light-weight file fragment classification based on convolutional neural network, called Depthwise Separable Convolutional (DSC). In this paper, we propose two additional models: Depthwise Separable Convolutional with Squeeze-and-Excitation (DSC-SE) and Modified Depthwise Separable Convolutional (M-DSC). We show that the proposed models reduce the number of parameters as compared to CNN. Therefore, the process of file fragment classification can be carried out without the need of specialized hardware, e.g., GPUs and TPUs. To evaluate the performance of our models, we compared the results of the proposed models with FiFTy \cite{mittal2020fifty} and a baseline Recurrent Neural Networks (RNN) \cite{hochreiter1997long}. We have trained the models using FFT-75 dataset \cite{fft-75}, which includes six scenarios. The results of Scenario 1 show that our model achieves an accuracy that is comparable to FiFTy using 6.3x less floating-point operations (FLOPs) on 4096 bytes fragment and 87x less FLOPs on 512 bytes fragments. Furthermore, our models achieve 79.27\% accuracy with around 100K parameters and 164M FLOPs, while FiFTy achieves 77.04\% accuracy with more than 400K parameters and 1 GFLOPs. Finally, we show that our models outperform FiFty in terms of inference time (milliseconds per block) by being 15x faster on GPU than FiFTy and 105x faster than baseline RNN. 

The rest of this paper is organized as follows. In Section 2, we provide the necessary background for this work. In Section 3, we present our proposed models. In Section 4, we discuss the experiments and evaluation results. In Section 5, we discuss the state-of-the-art of file fragment classification. Finally, we draw conclusions from our findings and suggest potential directions for future research in Section 6. 

\section{Preliminaries}
\subsection{File carving}
File carving is the process of extracting files from storage media (such as hard drives) based on their content instead of using the API or meta-data of a file system \cite{lin2018file}. Such process is often used when the underlying file system is damaged or unknown. File carving can be complicated when files are fragmented by the operating system. Therefore, file carving process includes carving file fragments, identifying each fragment's type, ordering the fragments, and assembling the file. There are several file carving tools and techniques that are used in practice, including header/footer matching and file structure-based carving\cite{poisel2013comprehensive,povar2010forensic}. However, there are few challenges that need to be addressed: 1) improving the performance of file carving, as the average digital forensics case size nowadays is in terabytes, and 2) improving the overall accuracy of file carving.

\subsection{Convolution Neural Network}
CNN is a type of artificial neural network that is widely used for object detection and  image classification. CNN contains one or more convolution layers that can extract features from the inputs and  predict an output based on the extracted features~\cite{fukushima1980self}. A convolution filter, also known as a kernel, is a 2D matrix that is used to perform the convolution operation on images. Mathematically, the convolution operation is the dot product between the kernel and the layer's input. Various kernels can be used to extract various features from the input data points. In general, kernel filters are designed to be smaller than the input to allow for the sharing of kernel weights across input dimensions. In file fragment classification, the input to the CNN is either a vector of 4096 or 512 dimensions, equivalent to fragments of 4KB or 512 bytes, respectively. The input is then transmitted to an embedding layer that converts the input vector into a continuous (4096,32) or (512,32) dimensional vector, dependent on the fragment size. The embedding layer is a preliminary step to prepare the input vectors for subseque CNN layers. The output of the embedding layer is fed into a convolution layer with kernels of any size. Several convolution layers can be stacked \cite{mittal2020fifty}. As a result of stacking standard convolution layers when developing a Deep Neural Network (DNN), in which the model can have a large number of parameters.
\subsection{Depthwise Separable Convolution}
\begin{figure}[t!]
\centering
        \begin{subfigure}[h!]{0.3\textwidth}
            \includegraphics[width=\textwidth,height=0.2\textheight]{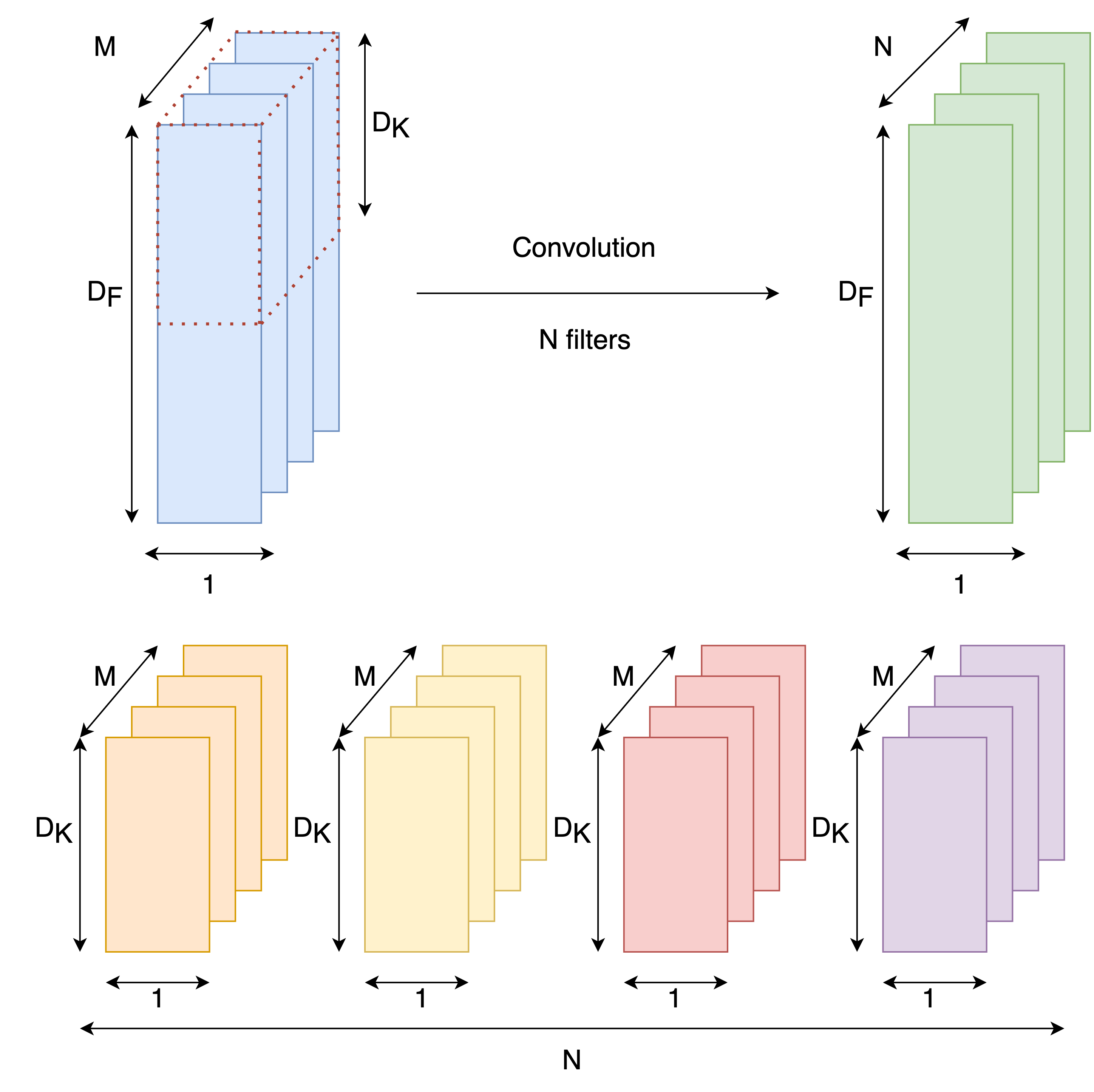}
            \caption{Standard Convolution}
            \label{fig:standard_convultion}
        \end{subfigure}
        \hfil
        \begin{subfigure}[H]{0.3\textwidth}
            \includegraphics[width=\textwidth,height=0.2\textheight]{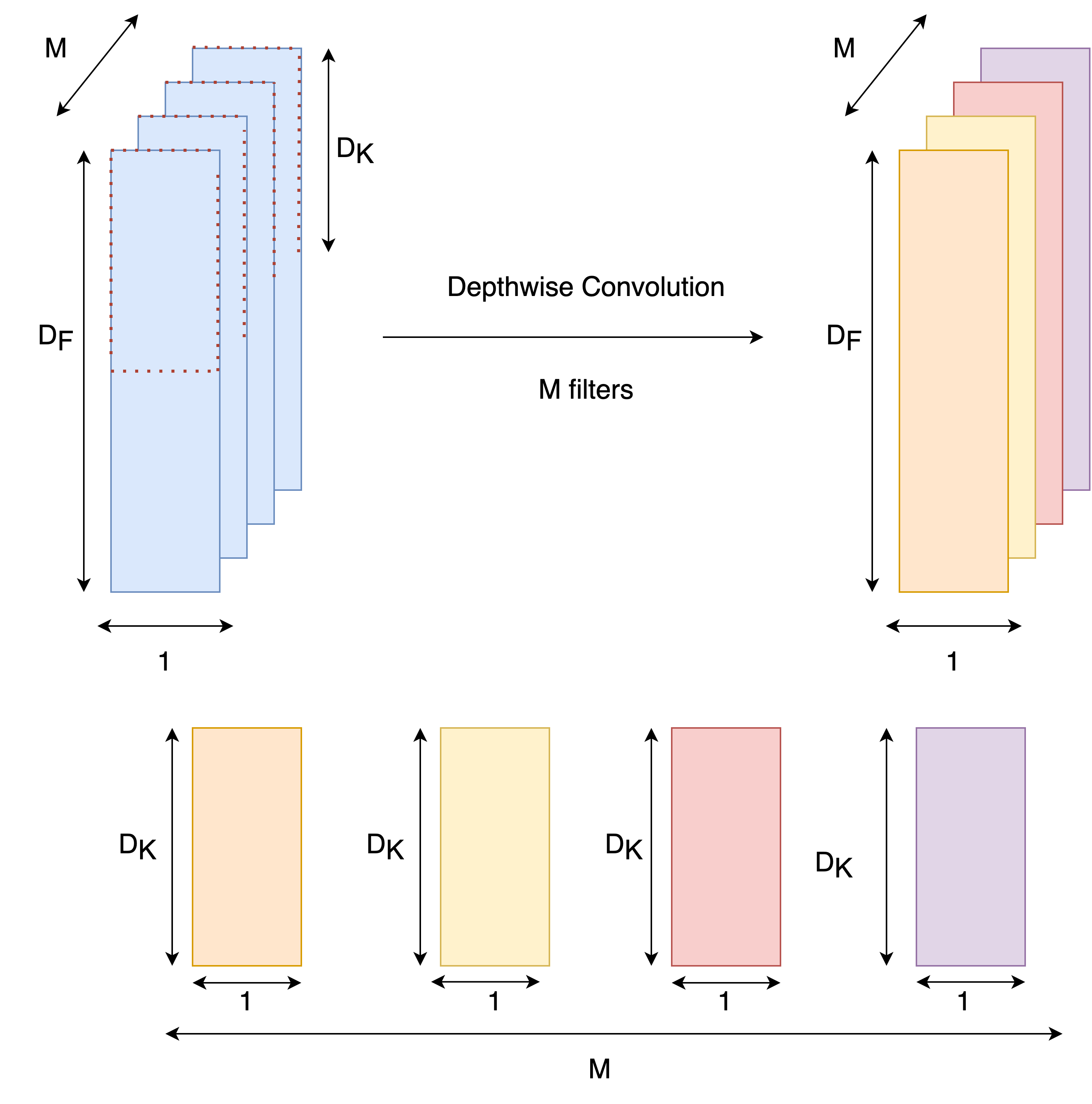}
            \caption{Depthwise Convolution}
            \label{fig:depthwise_convolution}
        \end{subfigure}
        \hfil
        \begin{subfigure}[H]{0.3\textwidth}
            \includegraphics[width=\textwidth,height=0.2\textheight]{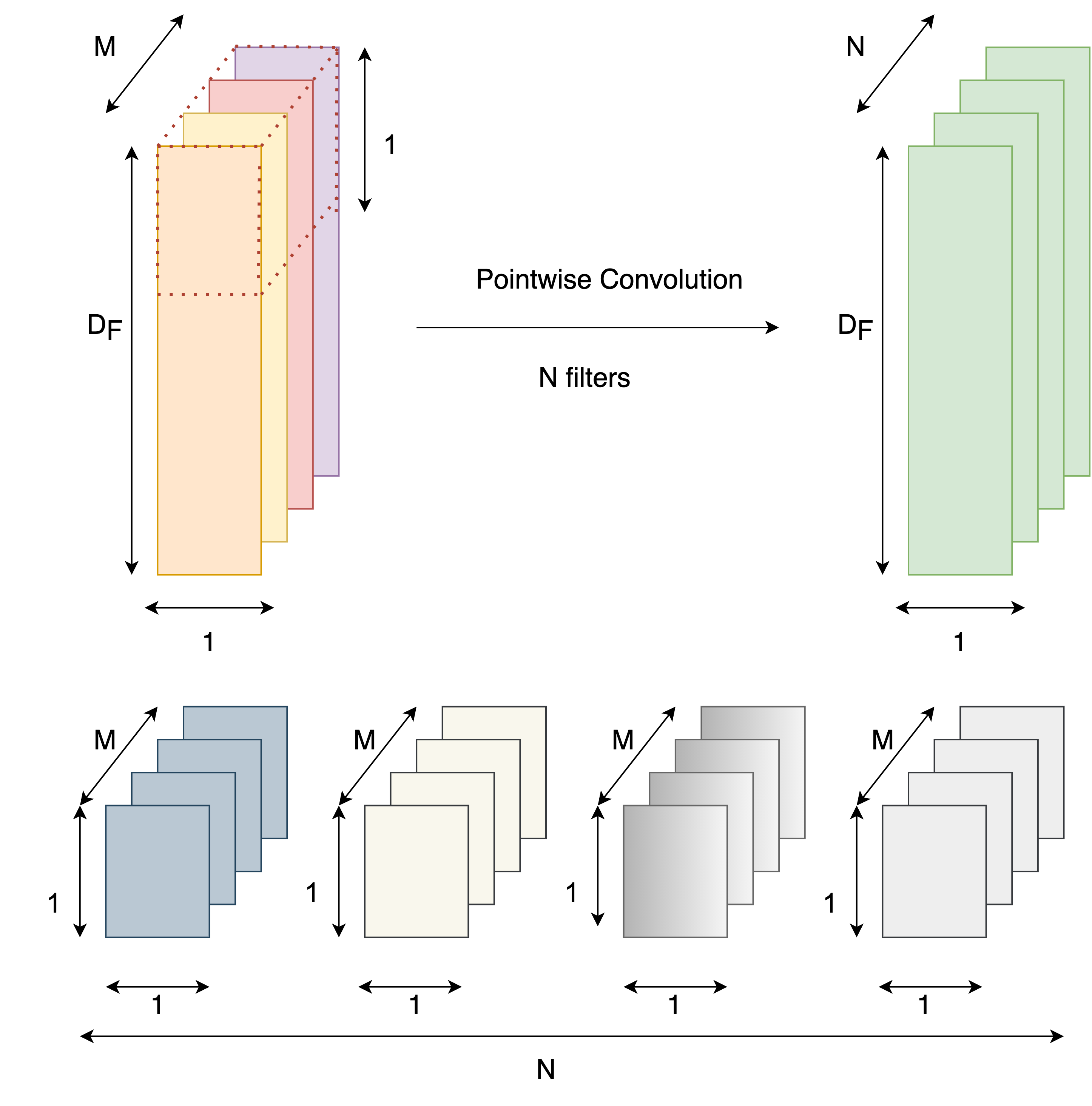}
            \caption{Pointwise Convolution}
            \label{fig:pointwise_convolution}
        \end{subfigure}
        \caption{Standard convolution in (a) is factorized into two layers: Depthwise Convolution in (b) and pointwise in (c).}
    \end{figure}
The separable convolution model was first introduced by Sifre and Mallat \cite{rigid_motion}. Since then, several models have been proposed based on separable convolution, including Xception-net \cite{xception}, Mobilenets \cite{mobilenet}, Effecienet \cite{sze2017efficient}, and Shufflenet \cite{zhang2018shufflenet}. The automatic feature extraction behaviour of standard convolution is achieved by smaller and faster depthwise separable convolution. Depthwise separable convolution is a form of factorized convolutions in which the normal convolution kernel is split into two parts: one depthwise convolution and a $1\times1$ convolution, called pointwise convolution. The splitting of convolution into two layers drastically reduce the number of parameters, and hence the computation time required for training and inference. The depthwise convolution applies a single filter to each input channel. The pointwise convolution then applies a $1\times1$ convolution to combine the outputs of the depthwise convolution \cite{mobilenet}. 
Figure \ref{fig:standard_convultion} depicts a standard one-dimensional convolution operation that takes $D\textsubscript{F}\times1\times M$ feature maps as input and produces $D\textsubscript{F}\times1\times N$ feature maps, where \textit{D\textsubscript{F}} is length of the input feature, \textit{M} is the number of input channels, and \textit{N} is the number of output channels. In convolutional network, the number of parameters with kernel size \textit{K} is $D\textsubscript{K}\times1\times N\times M$. Subsequently, the total computation cost is   \begin{equation}
     D\textsubscript{F}\cdot N\cdot M\cdot D\textsubscript{K}
    \end{equation}
    Unlike CNNs, depthwise separable convolution factorizes the convolution layer into two layers: depthwise convolution depicted in figure \ref{fig:depthwise_convolution} and pointwise convolution shown in figure \ref{fig:pointwise_convolution}. The depthwise convolution layer with one filter per input channel has $D\textsubscript{K}\times1\times M$ parameters with K sized kernel and M number of channels. Therefore, depthwise convolution has a computation cost of: 
    \begin{equation}
        D\textsubscript{F}\cdot M\cdot D\textsubscript{K}
    \end{equation}
    On the other hand, pointwise convolution uses a linear combination of filtered feature maps with the help of $1\times1$ convolution. The cost of $1\times1$ convolution is the same as standard convolution having a kernel size of 1, i.e., $D\textsubscript{F}\times N \times M$. Consequently, the total computation cost of depthwise separable convolution is:
    \begin{equation}
        D_{F}\cdot M\cdot D_{K} + D_{F}\cdot N\cdot M
    \end{equation}
    As a result, the total reduction in the computation cost when using depthwise separable convolution is:
    \begin{equation}
       \scalebox{1.3}{$ \frac{D_{F}\cdot M\cdot D_{K} + D_{F}\cdot N\cdot M}{D_{F}\cdot N\cdot M\cdot D_{K}}$}
    \end{equation}
    
    \begin{equation}
          = \scalebox{1.3}{$\frac{1}{N} + \frac{1}{D_{K}}$}
    \end{equation}
    
    The value of N ranges from 32 to 1024 and D\textsubscript{K} for one-dimensional convolution can be between 9 and 27. Therefore, the reduction in computation cost is between 85\% and 95\%.
    

    In this regard, several models have been created using depthwise separable convolution for the purpose of image classification. Such models achieve high accuracy while still enabling speedy inference time. \cite{mobilenet,mobilenetv2,xception}.

\section{Light-weight CNNs for File Fragment Classification}
     \begin{figure*}[]
        \centering
        \includegraphics[scale=0.2]{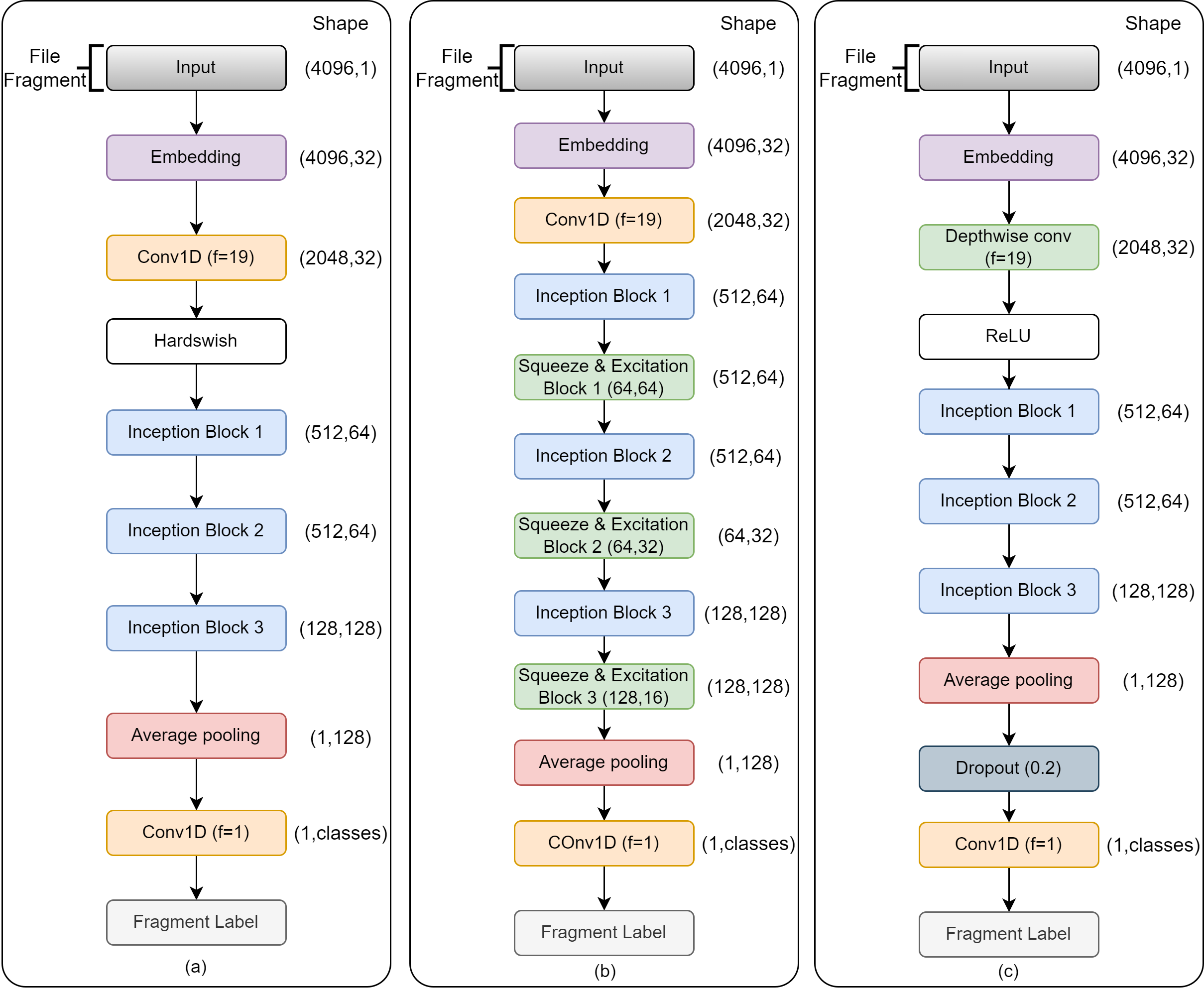}
        \caption{Network Architectures. a) DSC, b) DSC-SE, and c) M-DSC }
        \label{fig:Network arc}
    \end{figure*}
    In this section, we describe the proposed architectures of the proposed light-weight CNNs models for file fragment classification. Previous file carving techniques mostly rely on statistical feature extraction \cite{Karresand2006OscarF}, SVMs \cite{beebe2013sceadan}, CNN \cite{mittal2020fifty}, RNN \cite{Hiester2018FileFC}, and Fully Connected Neural Networks (FCNN) \cite{vulinovic2019neural}. RNNs are exponentially slow, while FCNNs and CNNs require a deep model to achieve good performance. Our objective is to reduce the inference time of file fragment classification while maintaining high accuracy. In this regard, several models have been developed based on depthwise separable convolution for image classification tasks. These models can achieve fast inference time while maintaining high accuracy \cite{mobilenet,mobilenetv2,xception}. Compared to FCNNs, CNNs, and RNNs, it was shown that depthwise separable convolution typically requires less computation time and fewer parameters \cite{mobilenet}.

  Figure \ref{fig:Network arc} illustrates the overall architectures of the proposed models. All models consist of an embedding layer and a standard 1D convolution block. The embedding layer transforms a file fragment byte ranging from 0 to 255 into a dense continuous vector of 32 dimensions. The embedding layers is used to compress the input feature dimension into a smaller space. Without embedding layer, each  byte value would be represented by a 256-dimensional sparse vector. With multiple layers of standard and depthwise separable convolutions, features are automatically derived by non-linearly transforming inputs. In order to extract essential features automatically for classification, the dense vector representation is then passed to a standard 1D convolution block and multiple depthwise separable inception blocks or squeeze-and-excitation blocks. To get the final features for classification, the output of the last depthwise separable convolution block is averaged along the spatial dimension. In the following sections, we provide more details about each model. 

  \subsection{Depthwise Separable Convolution (DSC)}
In DSC, the output of the 1D convolutional layer is passed to a hardswish activation function, which adds non-linearity to the model. The Hardswish function \cite{howard2019searching} is as the following. 
    \begin{equation}
    Hardswish(x) =
    \begin{cases} 
      0 & if x\leq -3, \\
      x & if x\geq +3,  \\
     \frac{x \cdot(x+3)}{6}  & otherwise
  \end{cases}
    \end{equation} 
The output is then processed by three inception blocks, which are composed of multiple parallel convolutional and pooling layers, designed to capture both local and global information from the input. After the inception blocks, the output is passed to an average pooling layer to reduce the spatial dimension, followed by a final 1D convolutional layer that performs feature map reduction and generates the final prediction. The class probabilities are determined from feature vectors using $1\times1$ convolution followed by \textit{softmax} activation function \cite{dunne1997pairing}.

\begin{figure}[]
        \centering
        \includegraphics[width=0.4\textwidth]{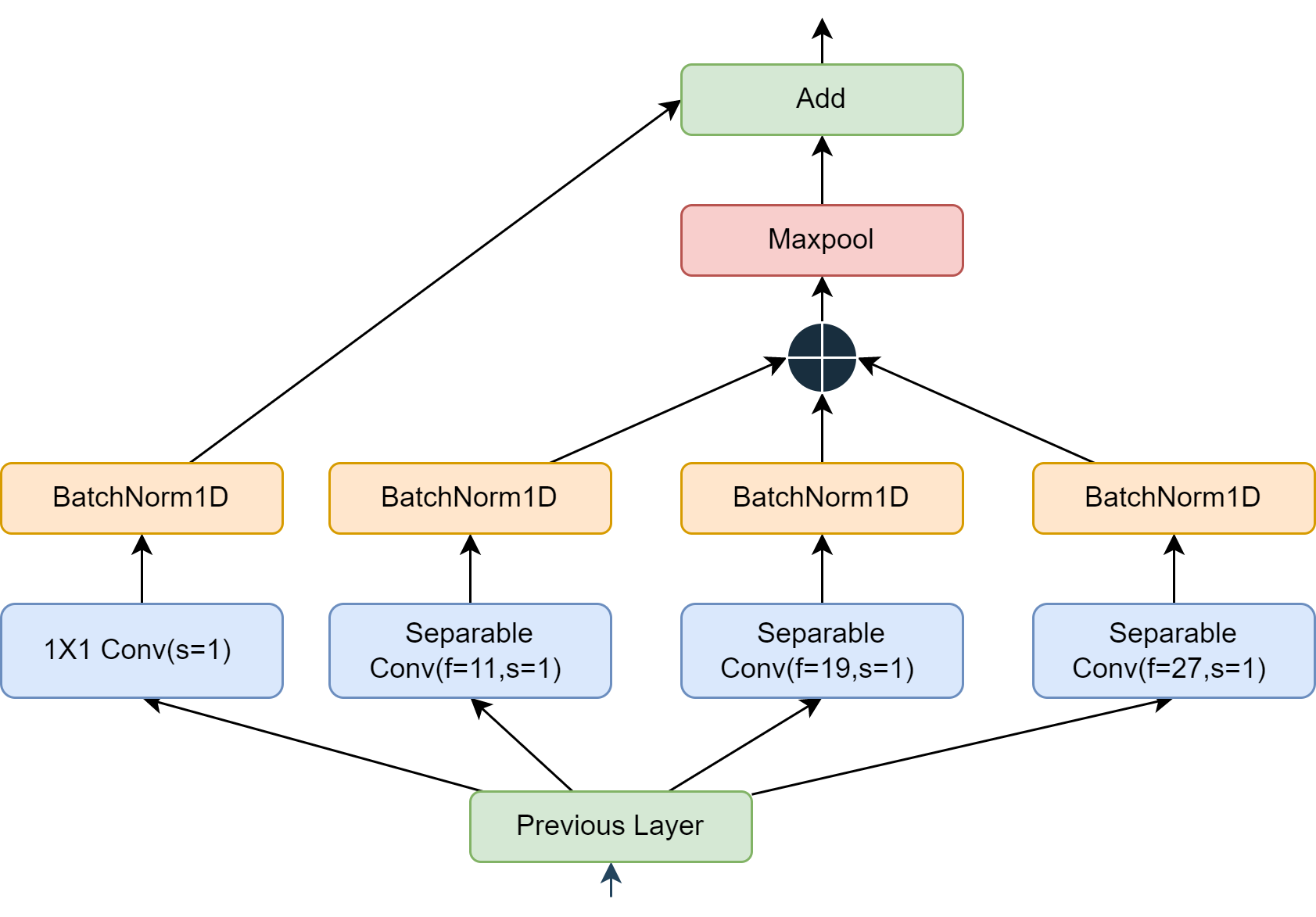}
        \caption{Inception Block}
        \label{fig:separable_conv}
    \end{figure}

\textbf{Inception Block.} Figure \ref{fig:separable_conv} depicts the architecture of a typical inception block. The previous layer's output is passed to three parallel depthwise separable convolutions and one $1\times1$ convolution. In our model, the kernel size of the depthwise separable convolution layer is 11, 19, and 27, respectively, with strides of 1, 1, 1, and 4, respectively. Kernel sizes were chosen empirically based on previous state-of-the-art network \cite{mittal2020fifty}. A batch normalization layer follows each of the convolution layer. All the depthwise separable convolution layer output are summed and then sub-sampled using a max-pooling layer with size 4. The output of the $1\times1$ convolution branch is directly added to the output of the max-pooling layer. If the number of the input and output channels is the same, then the $1\times1$ convolution layer is replaced with the direct addition of the output of max-pooling layer.   
        \begin{figure}[]
        \centering
        \includegraphics[width=0.4\textwidth]{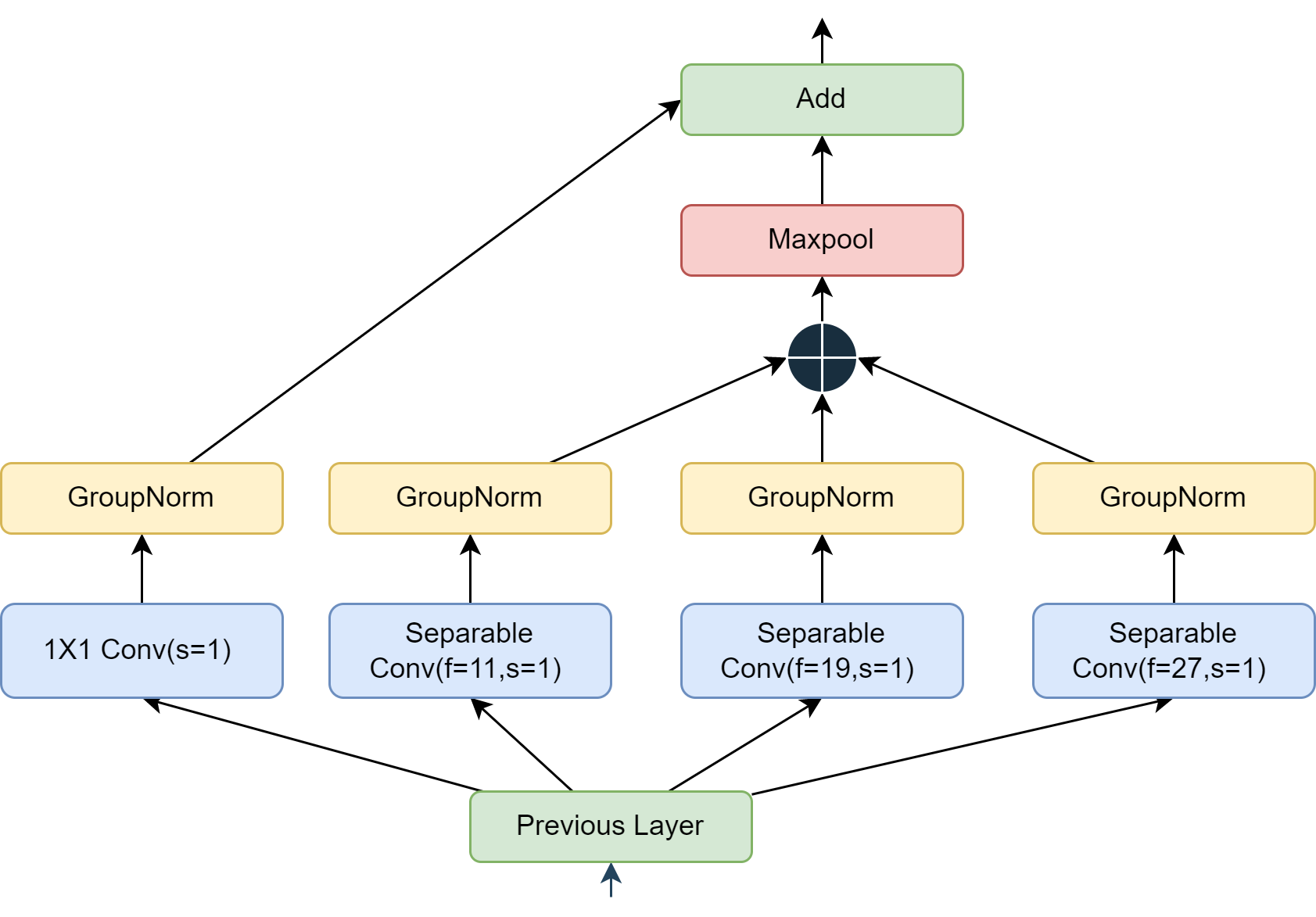}
        \caption{Modified Inception Block}
        \label{fig:modified_separable_conv}
    \end{figure}
    
        \begin{figure}[]
        \centering
        \includegraphics[width=0.2\textwidth]{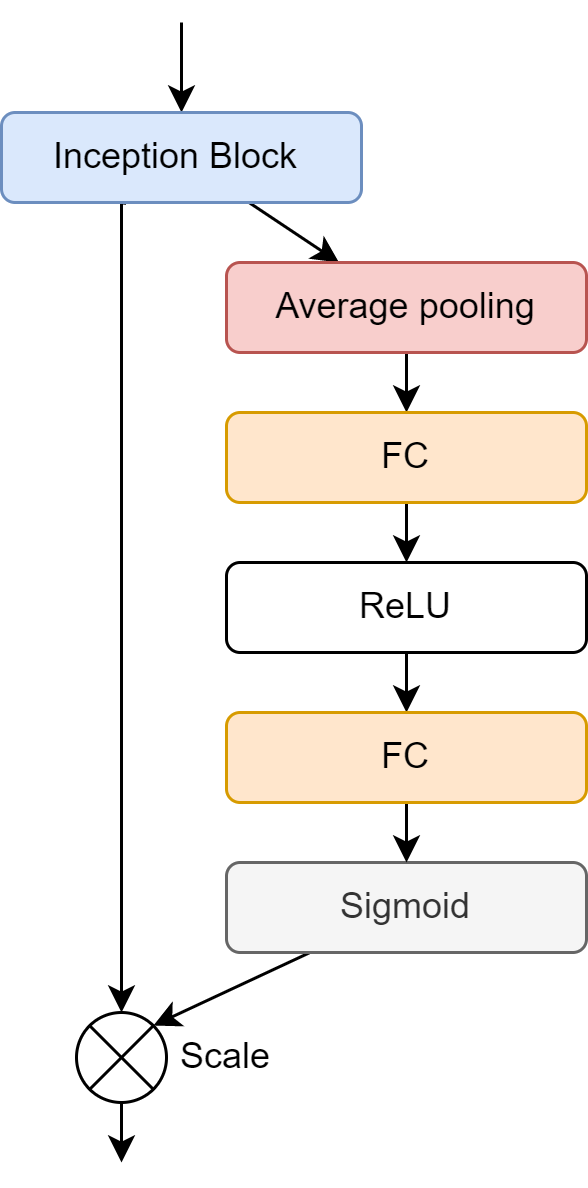}
        \caption{Squeeze-and-Excitation Block}
        \label{fig:sqeeze}
    \end{figure}

\subsection{Depthwise Separable Convolution with Squeeze and Excitation (DSC-SE)}
 DSC-SE is based on DSC but with the addition of Squeeze-and-Excitation (SE) \cite{hu2018squeeze} blocks after each inception block. SE block is a type of attention mechanism to improve CNNs performance. SE block allows the network to focus on the most important features in each channel of the feature maps, which is useful in tasks such as image classification, segmentation, and object detection. SE block works by first squeezing the spatial information from the feature maps into a single channel-wise global descriptor, and then using a fully connected layer to calculate a set of channel-wise weights. These weights are then multiplied element-wise with the feature maps. This, in turn, assigns higher weights to the channels that contain the most important information for the task. Recently, SE blocks have been integrated into the inception architecture resulting in an improved performance in various tasks~\cite{hu2018squeeze}. The core principle behind SE blocks is to give greater importance to the features that contribute more to the final prediction. Moreover, SE blocks are computationally efficient and easy to implement. Figure \ref{fig:sqeeze} shows the structure of the SE block in detail. 
  
 \subsection{Modified Depthwise Separable Convolution (M-DSC)}
 M-DSC architecture is also based on DSC but with several modifications. Instead of the first 1D convolutional layer, M-DSC uses a depthwise convolutional layer that performs convolutions independently on each channel of the input. The hardswish activation is replaced with the rectified linear unit (ReLU). Instead of batch normalization, group normalization was used to normalize the activation of multiple channels at once \cite{ioffe2015batch,wu2018group}. The modified inception block  used in our proposed models is shown in Figure \ref{fig:modified_separable_conv}. This results in reducing the memory and computation overhead of batch normalization. Finally, after the average pooling and before the final 1D convolutional layer, a dropout layer is added to prevent overfitting by randomly setting a fraction of the activations to zero during training. 
 
 In summary, the three architectures differ in the way of handling activation functions, normalization, and feature recalibration, with the aim to improve the performance and efficiency of the model. DSC-SE adds SE blocks to the inception blocks to improve feature recalibration, while M-DSC replaces the 1D convolutional layer with a depthwise convolutional layer, replaces hardswish with ReLU, and introduces group normalization and dropout to improve performance and efficiency.

\section{Performance Evaluation}
In this section, we present the results of evaluating the performance of our proposed model against baseline models based on RNN and FiFTy \cite{mittal2020fifty}. 

 \subsection{Dataset}
\begin{table}[t!]
   \caption{Grouping of different file types}
    \centering
    \begin{tabular}{|c|l|}
    \hline
        \textbf{Grouping} & \textbf{Files} \\
        \hline
        Archive & apk, jar, msi, dmg, 7z, bz2, deb, gz,\\ &  pkg, rar, rpm, xz, zip \\
        \hline
        Audio & aiff, flac, m4a, mp3, ogg, wav, wma \\
        \hline
        Bitmap & jpg, tiff, heic, bmp, gif, png \\
        \hline
        Executable & exe, mach-o, elf, dll \\
        \hline
        Human-readable & md, rtf, txt, tex, json, html, xml, log,\\ & csv\\
        \hline
        Office & doc, docx, key, ppt, pptx, xls, xlsx \\
        \hline
        Published & djvu, epub, mobi, pdf \\
        \hline
        Raw & arw, cr2, dng, gpr, nef, nrw, orf, pef, raf,\\& rw2, 3fr \\
        \hline
        Vector & ai, eps, psd \\
        \hline
        Video & mov, mp4, 3gp, avi, mkv, ogv, webm \\
        \hline
        Miscellaneous & pcap, ttf, dwg, sqlite \\
        \hline
    \end{tabular}
    \label{tab:grouping}
\end{table}
    To evaluate the performance of our models, we used the dataset provided by Mittal et al. in \cite{fft-75}, which contains a balanced number of files per class. Other datasets, e.g., \cite{garfinkel_farrell_roussev_dinolt_2009}, are highly imbalanced with 20 files-types comprising 99.3\% of the dataset and remaining 0.7\% belonging to 43 file types. The dataset used is composed of 75 types of files that are organized into 6 different scenarios. The Scenarios are described as follows in \cite{mittal2020fifty}:
    \begin{enumerate}
        \item (All; 75 classes): All file types are separate classes; this is the most generic case and can be aggregated into more specialized use-cases.
        \item (Use-specific; 11 classes): File types are grouped into 11 classes according to their use; this information may be useful for elaborate, hierarchical classification, or for determining the primary use of an unknown device.
        \item (Media Carver - Photos \& Videos; 25 classes): Every file type tagged as a bitmap (6), RAW photo (11), or video (7) is considered as a separate class; all remaining types are grouped into one other class.
        \item (Coarse Photo Carver; 5 classes): Separate classes for different photographic types: JPEG, 11 RAW images, 7 videos, 5 remaining bitmaps are grouped into one separate class per category; all remaining types are grouped into one other class.
        \item (Specialized JPEG Carver; 2 classes): JPEG is a separate class and the remaining 74 file types are grouped into one other class; this scenario is designed for analyzing disk images from generic devices.
        \item (Camera-Specialized JPEG Carver; 2 classes): JPEG is a separate class, and the remaining photographic/video types (11 RAW images, 3GP, MOV, MKV, TIFF and HEIC) are grouped into one other class; this scenario is designed for analyzing SD cards from digital cameras.
    \end{enumerate}
    \subsection{Baseline Model}
   We implemented a variation of RNN, i.e., Long Short Term Memory (LSTM) based model, as a baseline model for performance comparison based on the work in \cite{Hiester2018FileFC}. It was observed that the low number of learnable parameters hindered their effectiveness in classifying 75 types of files. The model specifications are as follows. File fragment byte sequences are fed into a 32-dimensional embedding layer followed by bidirectional and unidirectional LSTM layers. The LSTM layer has 128 neurons in each of its two layers for 512-byte fragments and 128 and 256 neurons in its bidirectional and unidirectional layers for 4 KB fragments, respectively. A softmax dense layer was used to produce the output labels.

    
\subsection{Experimental setup}
 \begin{table*}[]
    \caption{Comparison of results of five models on all 75 file types}
    \centering
    \begin{adjustbox}{max width=\textwidth}
    \begin{tabular}{l l l l l l l }
         \hline
         \textbf{Model} & \textbf{Neural Network} & \textbf{Block Size} & \textbf{\# Params} & \textbf{Accuracy} & \textbf{Speed[ms/block]$^{\dagger}$} & \textbf{Speed [min/GB]$^{\dagger}$}  \\
         \hline
         \multirow{2}{5em}{DSC} & \multirow{2}{10em}{Depthwise Separable CNN} & 4096 & 103,083 & 78.45 & \textbf{2.567} & \textbf{0.055} \\
         & & 512 & 103,083 & 65.89 & 3.531 & 0.382 \\
         \hline
        \multirow{2}{5em}{DSC-SE} & \multirow{2}{10em}{Depthwise Separable CNN with SE} & 4096 & 105,515 & \textbf{79.27} & 3.908 & 0.064 \\
         & & 512 & 105,515 & 66.33 & 3.473 & 0.471 \\
         \hline
        \multirow{2}{5em}{M-DSC} & \multirow{2}{10em}{Modified Depthwise Separable CNN} & 4096 & \textbf{85,291} & 78.68 & 2.606 & 0.057\\
         & & 512 & \textbf{85,291} & 64.03 & \textbf{2.776} & \textbf{0.378} \\
         \hline
         \multirow{2}{5em}{FiFTy} & \multirow{2}{10em}{1-D CNN} & 4096 & 449,867 & 77.04 & 38.189 & 1.366 \\
         & & 512 & 289,995 & 65.66 & 38.67 & 3.052 \\
         \hline
         \multirow{2}{5em}{Baseline RNN} & \multirow{2}{10em}{LSTM} & 4096 & 717,643 & 70.51 & 268.58 & 36.375 \\
         & & 512 & 379,851 & \textbf{67.5} & 126.54 & 33.431 \\
         \hline
         $^{\dagger}$ Computed on  Nvidia Titan X
    \end{tabular}
    \end{adjustbox}
    \label{tab:comp_all_3_75}
\end{table*}
    All of our experiments were conducted on a dual Intel Xeon CPU E5-2620 CPUs @ 2.40 GHz (12 physical cores, 24 logical cores), 192 GB RAM, and a single Nvidia Titan X GPU, running on Ubuntu 20.04 Operating System. Pytorch 1.5.0 was used for neural network design. We used automated hyper-parameter tuning for learning rate, optimizer choice, and activation function using TPE \cite{TPE} implemented through Optuna \cite{optuna_2019}. We found out that Adam optimizer \cite{kingma2014adam} and Hardswish \cite{howard2019searching} are the best optimizer and activation function, respectively. We did not perform tuning of convolution kernel size; we used the kernel sizes from previous work \cite{mittal2020fifty} that proved to be effective. The kernel sizes for different branches of the inception block were taken as 11,19, and 27. 
    
    The accuracy for different networks was calculated as follows:
        
        \begin{equation}
           Accuracy = \frac{TP + TN}{TP + TN + FP + FN} \\
        \end{equation}

    where TP is True Positive, TN is True Negative, FP is False Positive, and FN is False Negative.
    
    To achieve higher accuracy from smaller models, we pretrained our network on corresponding fragment size dataset. We tried to leverage the performance of transfer learning in gaining higher accuracy \cite{pratt1993discriminability}. In particular, to develop a 4096-byte fragment, the model was pretrained on 512-byte fragment dataset and vice-versa. We found that 6-8\% accuracy was increased when pretraining was done using 512-byte fragment data for developing the 4096-byte fragment model. However, similar performance was not achieved when pretraining was done 4096-byte fragment data for the 512-byte fragment model.

\subsection{Results}
\begin{table*}[t]
    \caption{Comparison between FiFTy and our models in terms of inference time and model Parameters}
    \centering
      \begin{adjustbox}{max width=\textwidth}
    \begin{tabular}{|c | l | l l l l| l  l l l| l l l l| }
         \hline
         \textbf{Scenario} & \textbf{Fragment} &   \multicolumn{4}{c|}{\textbf{\#param}}
          & \multicolumn{4}{c|}{\textbf{Inf. Time (CPU)[ms/block]}} & \multicolumn{4}{c|}{\textbf{Inf. Time (GPU)[ms/block]}}\\
         & \textbf{Size} & \textbf{DSC}  & \textbf{DSC-SE} & \textbf{M-DSC} & \textbf{FiFTy} & \textbf{DSC}  & \textbf{DSC-SE} & \textbf{M-DSC}& \textbf{FiFTy}& \textbf{DSC}  & \textbf{DSC-SE} & \textbf{M-DSC} & \textbf{FiFTy}\\
         \hline
         \multirow{2}{1em}{1} & 4096 & 103,083 & 105,515 & \textbf{85,291} & 449,867 & 18.595 & 23.291 & \textbf{17.338} & 121.476 & \textbf{2.567} & 3.908 & 2.606 & 38.189 \\
         & 512 & 103,083  & 105,515 & \textbf{85,291} & 289,995 & 6.958 & 8.216 & \textbf{6.279} & 75.551 & 3.531 & 3.473 & \textbf{2.776} & 38.673 \\
         \hline
         \multirow{2}{1em}{2} & 4096 & 94,827 & 97,259 & \textbf{77,035} & 597,259 & 22.587 & 18.023 & \textbf{17.214} & 92.324 & 2.763 & 3.300 & \textbf{2.610} & 29.826\\
         & 512 & 94,827  & 97,259 & \textbf{77,035} & 269,323 & 7.827 & 6.657 & \textbf{6.346} & 89.344 & 3.239 & 3.504 &\textbf{2.910} & 29.809 \\
         \hline
         \multirow{2}{1em}{3} & 4096 & 96,633 & 99,065 & \textbf{78,841} &  453,529 & \textbf{17.068} & 18.860 & 17.361 & 102.808 &3.284 & 3.182 & \textbf{2.632} & 37.286 \\
         & 512 & 96,633  & 99,065 & \textbf{78,841} & 690,073 & \textbf{6.066} & 6.613 & 6.300 & 99.710 & 3.078 & 3.499 & \textbf{2.708} & 35.361 \\
         \hline
         \multirow{2}{1em}{4} & 4096 & 94,053 & 96,485 & \textbf{76,261} & 684,485 & \textbf{17.149} & 20.477 & 17.537 & 100.117 & \textbf{2.524} & 3.197 & 2.608 & 40.176 \\
         & 512 & 94,053 & 96,485 & \textbf{76,261} & 474,885 & \textbf{6.029} & 6.533 & 6.285 & 79.346 & \textbf{2.620} & 3.258 & 2.677 & 35.965 \\ 
         \hline
         \multirow{2}{1em}{5} & 4096 & 93,666 & 96,098 & \textbf{75,847} & 138,386 & 22.427 & \textbf{16.362} & 17.474 & 99.855 & \textbf{2.524} & 3.362 & 2.769 & 39.262 \\
         & 512 &  93,666 & 96,098 & \textbf{75,847} & 336,770 & \textbf{5.967} & 6.327 & 6.275 & 79.831 & \textbf{2.581} & 3.259 & 2.873 & 42.413 \\
         \hline
         \multirow{2}{1em}{6} & 4096 & 93,666 & 96,098 & \textbf{75,847} & 666,242 & 18.418 & 23.000 & \textbf{17.279} & 98.536 & \textbf{2.530} & 3.389 & 2.641 & 34.931 \\
         & 512 & 93,666 & 96,098 & \textbf{75,847} & 242,114 & \textbf{6.133} & 8.373 & 6.398 & 89.104 & \textbf{2.612} & 3.274 & 2.702 & 38.982 \\
         \hline
    \end{tabular}
    \end{adjustbox}
    \label{tab:Comp_our_fifty_inf}
\end{table*}
In general, our models perform better than FiFTy in inference time with no significant loss of accuracy. The results are summarised in Table \ref{tab:comp_all_3_75}. In particular, we observe that our models outperform both FiFTy and RNN (Baseline) in terms of accuracy and inference time for scenario 1 (75 file types). In particular, FiFTy and RNN models are 25x and 660x, respectively, slower than our proposed models when running on 1 GB data of 4096-byte fragments. Using 512-byte fragments, FiFTy is 8x slower, and RNNs is 87x slower than our model when running on 1 GB of data. It was observed that RNN model performs well with 512-byte fragments due to their ability to handle short sequences. In contrast, when the fragment size increases to 4096-bytes, the performance deteriorates due to the vanishing gradient problem of RNNs \cite{hochreiter1998vanishing}. While RNN model has higher accuracy on the 512-byte fragments, RNN is not practical because it takes relatively longer inference time.

\textbf{Number of parameters and inference time.}
Table \ref{tab:Comp_our_fifty_inf} summarizes the number of neural network parameters in the proposed models compared to FiFTy. For all six scenarios and both fragment sizes, our models have far fewer parameters than FiFTy. In addition, Table \ref{tab:Comp_our_fifty_inf} compares the inference times for each of the six scenarios on both the GPU and CPU. Our models outperform FiFTy by a large margin on GPU and CPU in term of inference time for all six scenarios. We observed a 5x reduction in inference time for 4096-byte file fragments and a 9x reduction for 512-byte file fragments compared to FiFTy on CPU. For GPU, the time reduction is more than 9x for both fragment types.

\textbf{Number of FLOPS.}
To provide hardware-independent metrics, we calculated the floating-point operations in our models and FiFTy. In Table \ref{tab:Comp_our_fifty_flops}, we provide the results of the comparison. Compared to FiFTy, our models have 6.3x fewer FLOPS for 4096-byte fragments and 87x fewer for 512-byte fragments. In Scenario 1, for example, with a fragment size of 4096-bytes, DSC has 164.88, DSC-SE has 164.96, and M-DSC has 89.07 MFLOPs, while FiFTy has 1047.59 MFLOPs. These results demonstrate the effectiveness of our proposed models, as they require fewer computations to make predictions while still maintaining high accuracy. This makes our models more suitable for real-world applications where computational resources are limited. 


\begin{table}[]
    \caption{Comparison between FiFTy and our models for floating point operations (Mega FLOPs)}
    \centering
        \begin{adjustbox}{max width=0.47\textwidth}
    \begin{tabular}{|c | c | l l l l|}
        \hline
         \textbf{Scenario}  & \textbf{Fragment} & \multicolumn{4}{c|}{\textbf{FLOPs}}   \\
           & \textbf{size} & \textbf{DSC} & \textbf{DSC-SE} & \textbf{M-DSC} & \textbf{FiFTy} \\
        \hline
        \multirow{2}{1em}{1}  & 4096 & 164.88 & 164.96 & \textbf{89.07} & 1047.59 \\
        &  512 & 20.63 & 20.64& \textbf{11.15} & 1801.71 \\
        \hline
        \multirow{2}{1em}{2}  & 4096 & 164.86 & 164.95 & \textbf{89.05} & 1327.90 \\
        &  512 & 20.61 & 20.63 & \textbf{11.13} &  918.06 \\
        \hline
        \multirow{2}{1em}{3}  & 4096 & 164.86 & 164.95 &\textbf{89.05} & 647.78 \\
        &  512 & 20.61 & 20.63 & \textbf{11.14} &  3579.57 \\
        \hline
        \multirow{2}{1em}{4}  & 4096 & 164.86 &  164.95& \textbf{89.05}& 2378.51 \\
        &  512 & 20.61 & 20.63 & \textbf{11.13} & 1576.71 \\
        \hline 
        \multirow{2}{1em}{5}  & 4096 & 164.86 & 164.95 &\textbf{89.05} & 488.37 \\
        &  512 & 20.61 & 20.63 & \textbf{11.13} & 2330.48 \\
        \hline
        \multirow{2}{1em}{6}  & 4096 & 164.86 & 164.95 &\textbf{89.05} & 1126.00 \\
        &  512 & 20.61 & 20.63 &  \textbf{11.13} & 611.30 \\
        \hline
    \end{tabular}
        \end{adjustbox}
    \label{tab:Comp_our_fifty_flops}
\end{table}

\begin{figure*}
 \center
    \begin{subfigure}[h!]{0.30\textwidth}
                \includegraphics[width=\linewidth]{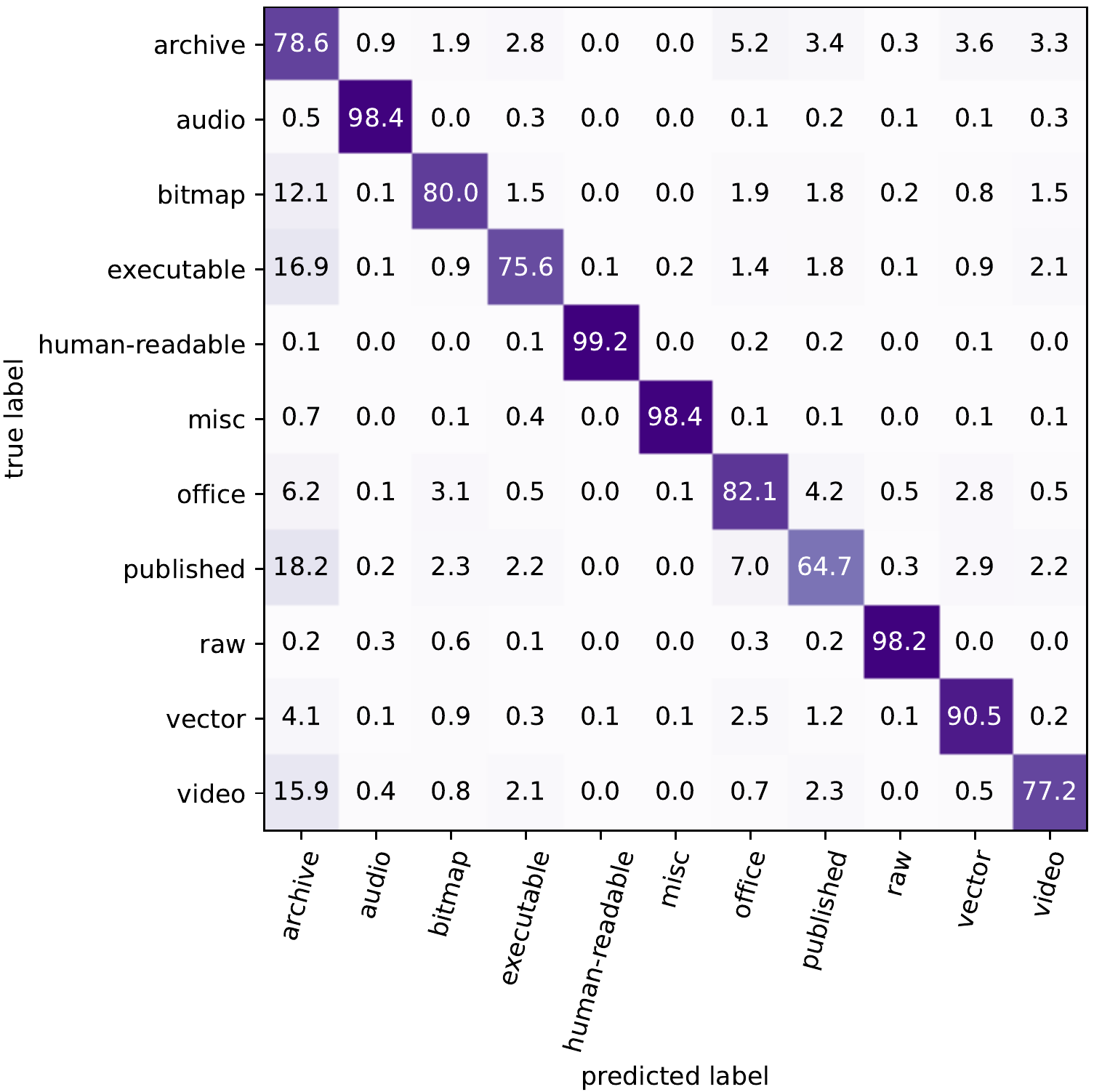}
                \caption{DSC (78.45\%)}
                \label{fig:ours_4K}
        \end{subfigure}%
                \begin{subfigure}[h]{0.30\textwidth}
                \includegraphics[width=\linewidth]{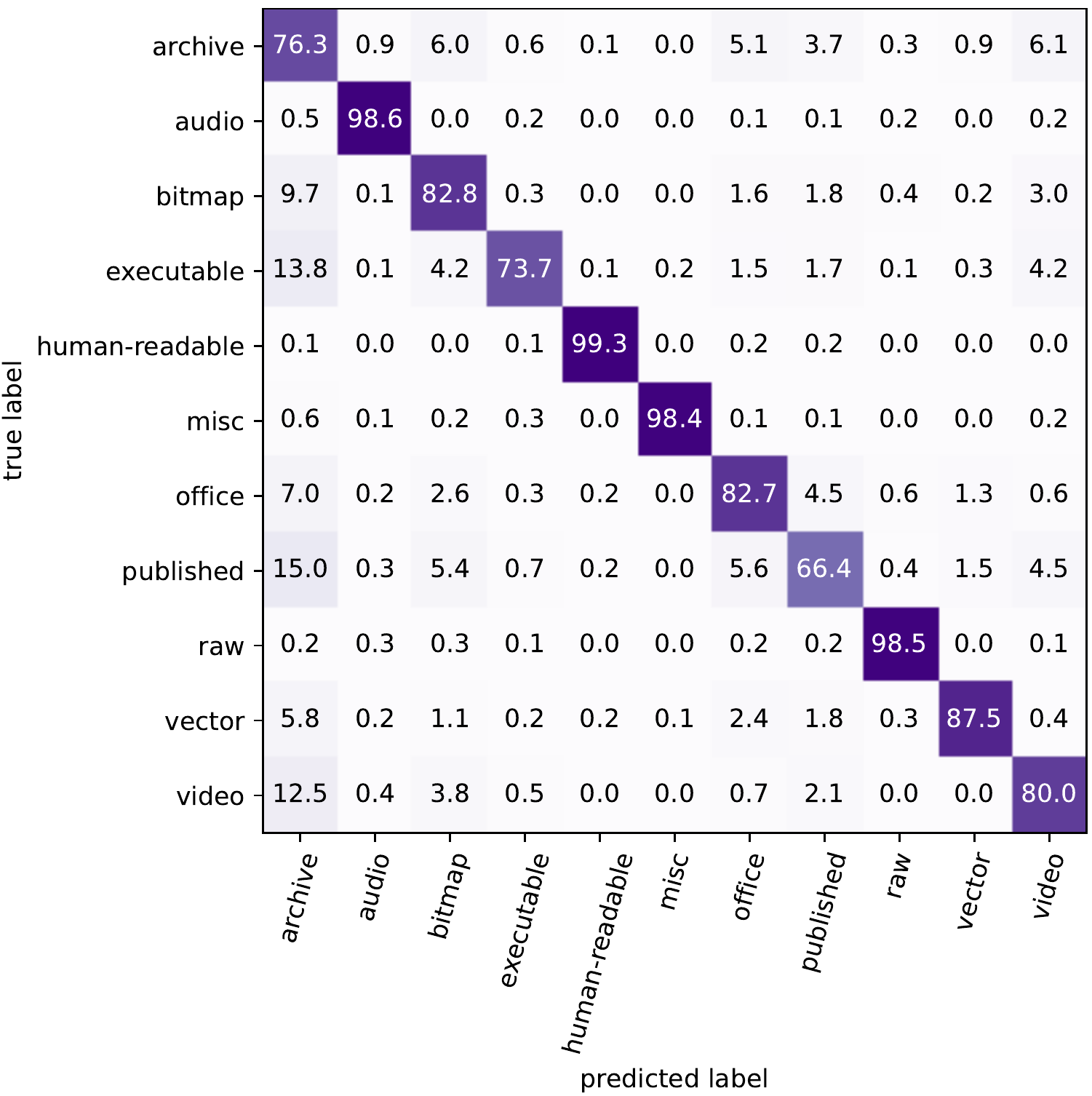}
                \caption{DSC-SE (79.27\%)}
                \label{fig:model2_4096}
        \end{subfigure}%
                \begin{subfigure}[h]{0.30\textwidth}
                \includegraphics[width=\linewidth]{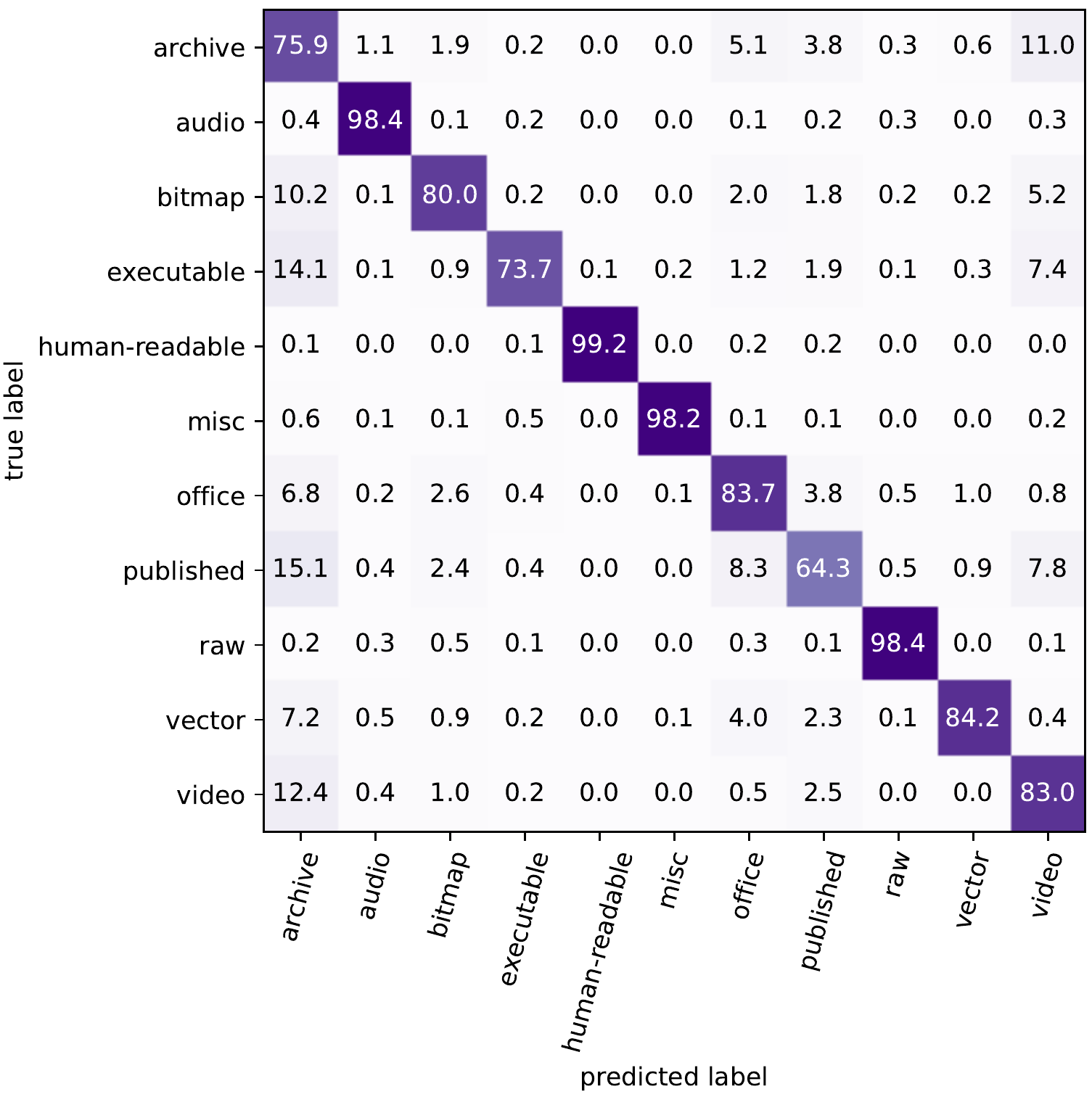}
                \caption{M-DSC (78.68\%)}
                \label{fig:model3_4096}
        \end{subfigure}%
        \linebreak
        \begin{subfigure}[h!]{0.30\textwidth}
                \includegraphics[width=\linewidth]{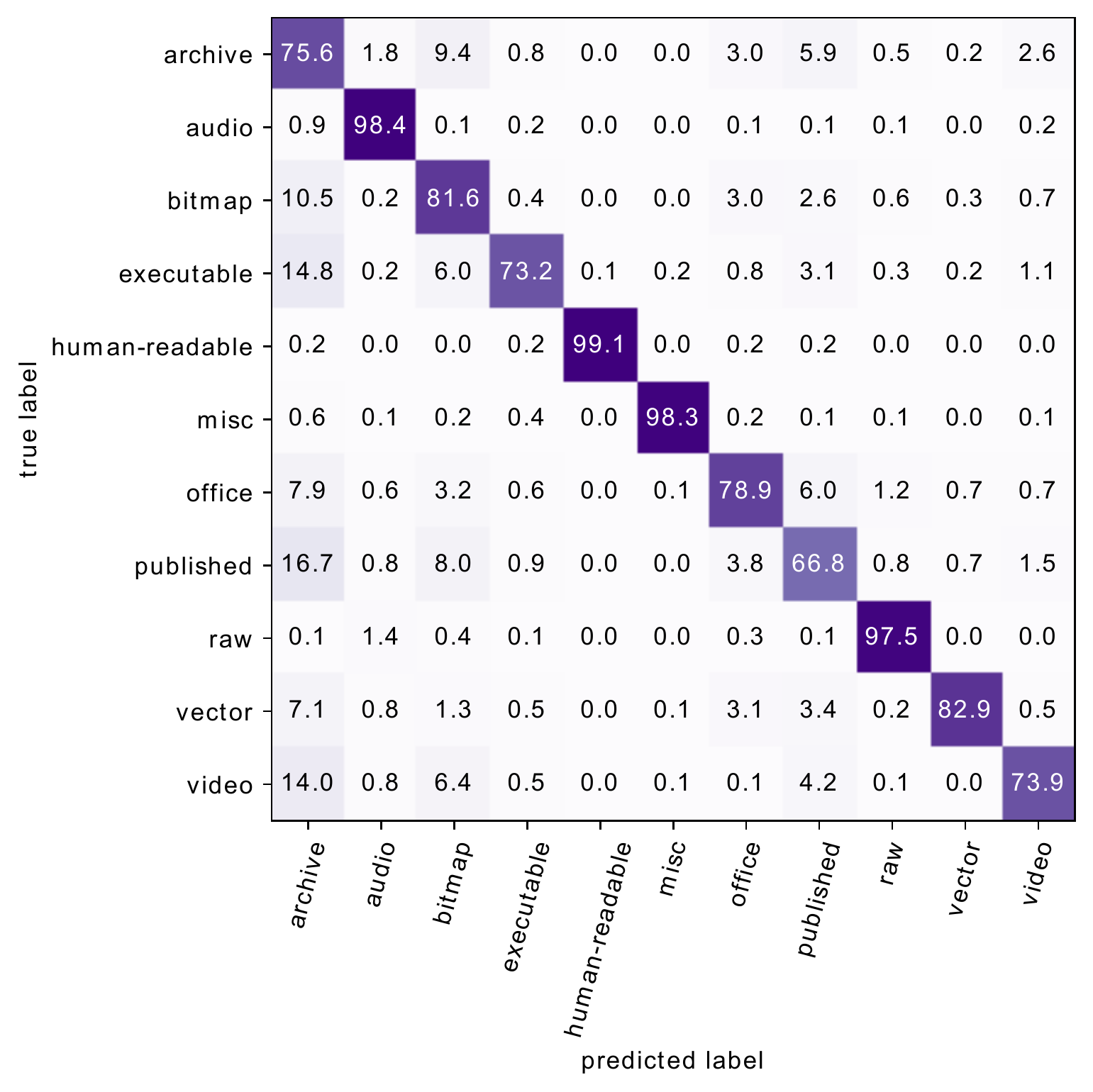}
                \caption{FiFTy (77.04\%)}
                \label{fig:fifty_4K}
        \end{subfigure}%
        \begin{subfigure}[h!]{0.30\textwidth}
                \includegraphics[width=\linewidth]{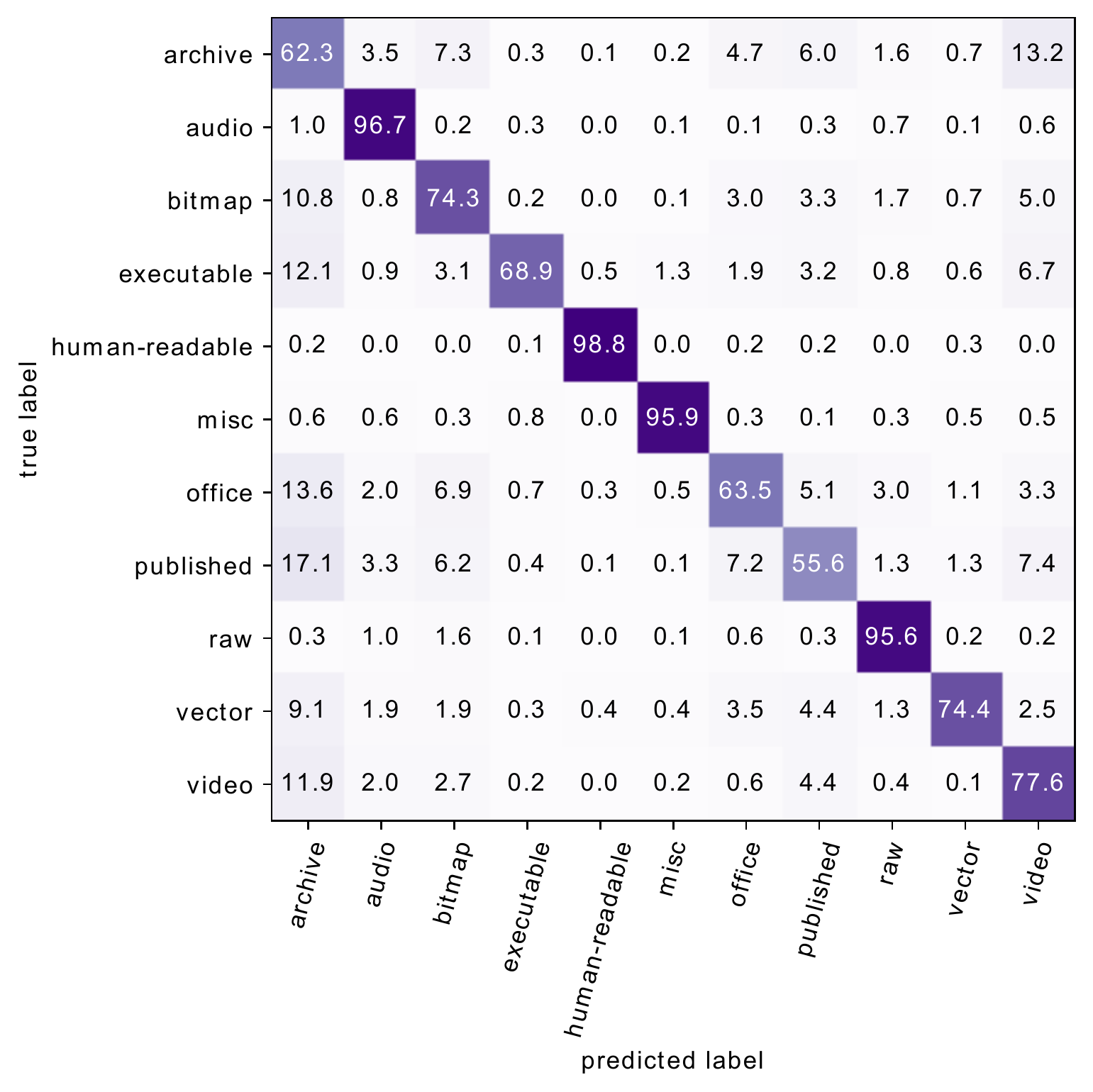}
                \caption{Baseline (70.51\%)}
                \label{fig:baseline_4K}
        \end{subfigure}%
        \linebreak
        \begin{subfigure}[h!]{0.30\textwidth}
                \includegraphics[width=\linewidth]{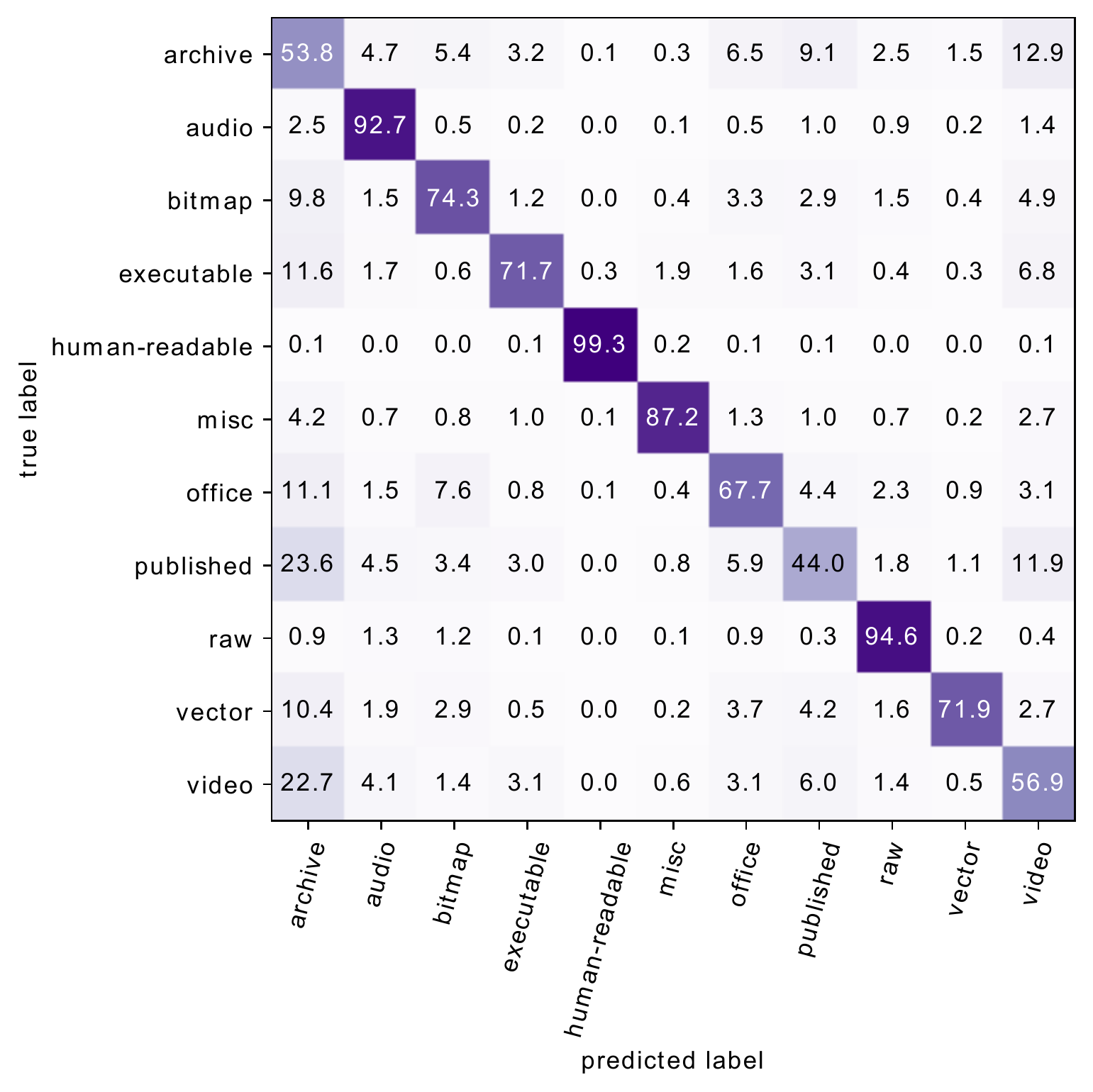}
                \caption{DSC (65.89\%)}
                \label{fig:ours_512}
        \end{subfigure}%
        \begin{subfigure}[h!]{0.30\textwidth}
                \includegraphics[width=\linewidth]{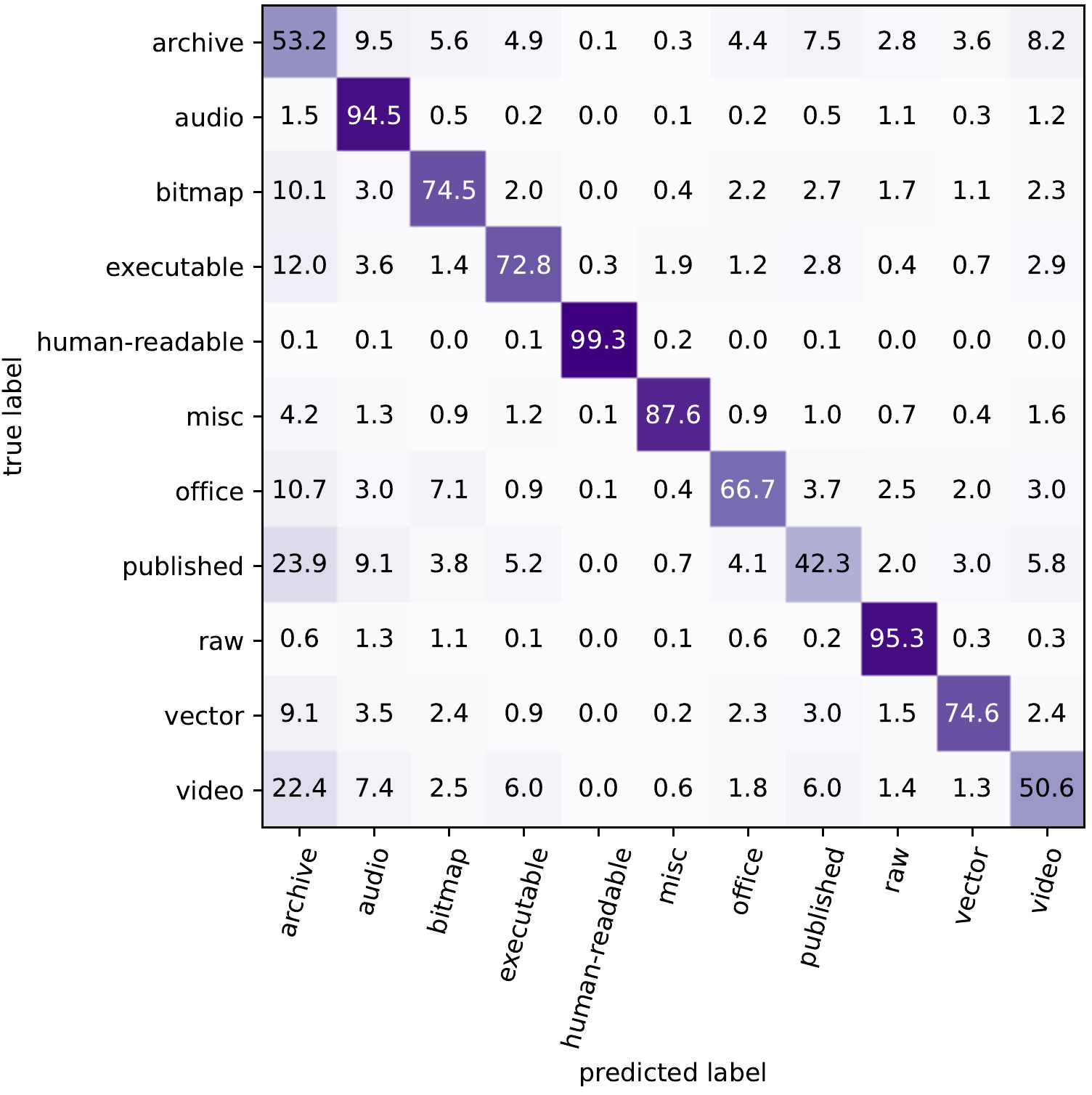}
                \caption{DSC-SE (66.33\%)}
                \label{fig:ours_512_2}
        \end{subfigure}%
        \begin{subfigure}[h!]{0.30\textwidth}
                \includegraphics[width=\linewidth]{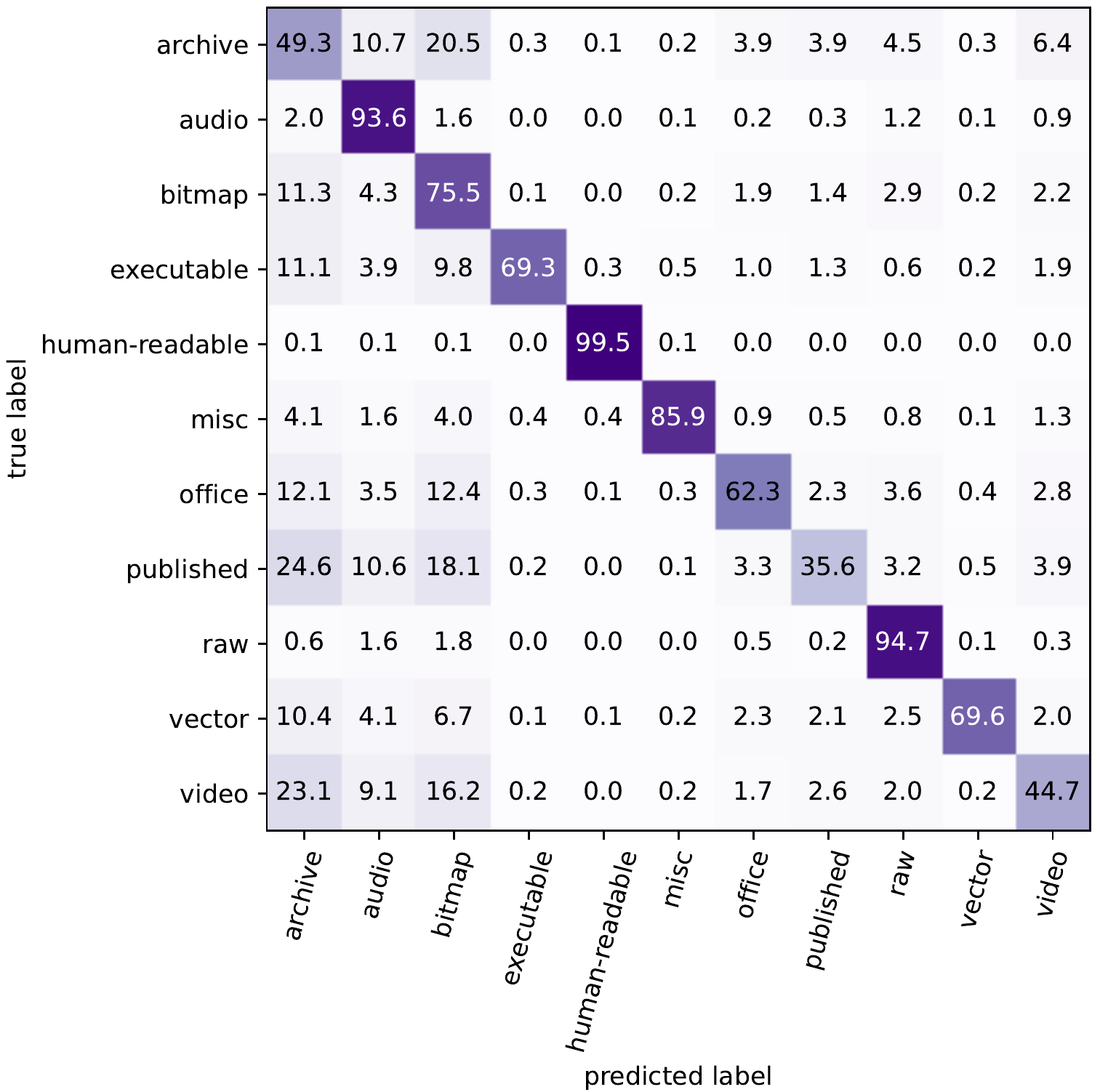}
                \caption{M-DSC (64.04\%)}
                \label{fig:ours_512_3}
        \end{subfigure}%
        \linebreak
        \begin{subfigure}[h!]{0.30\textwidth}
                \includegraphics[width=\linewidth]{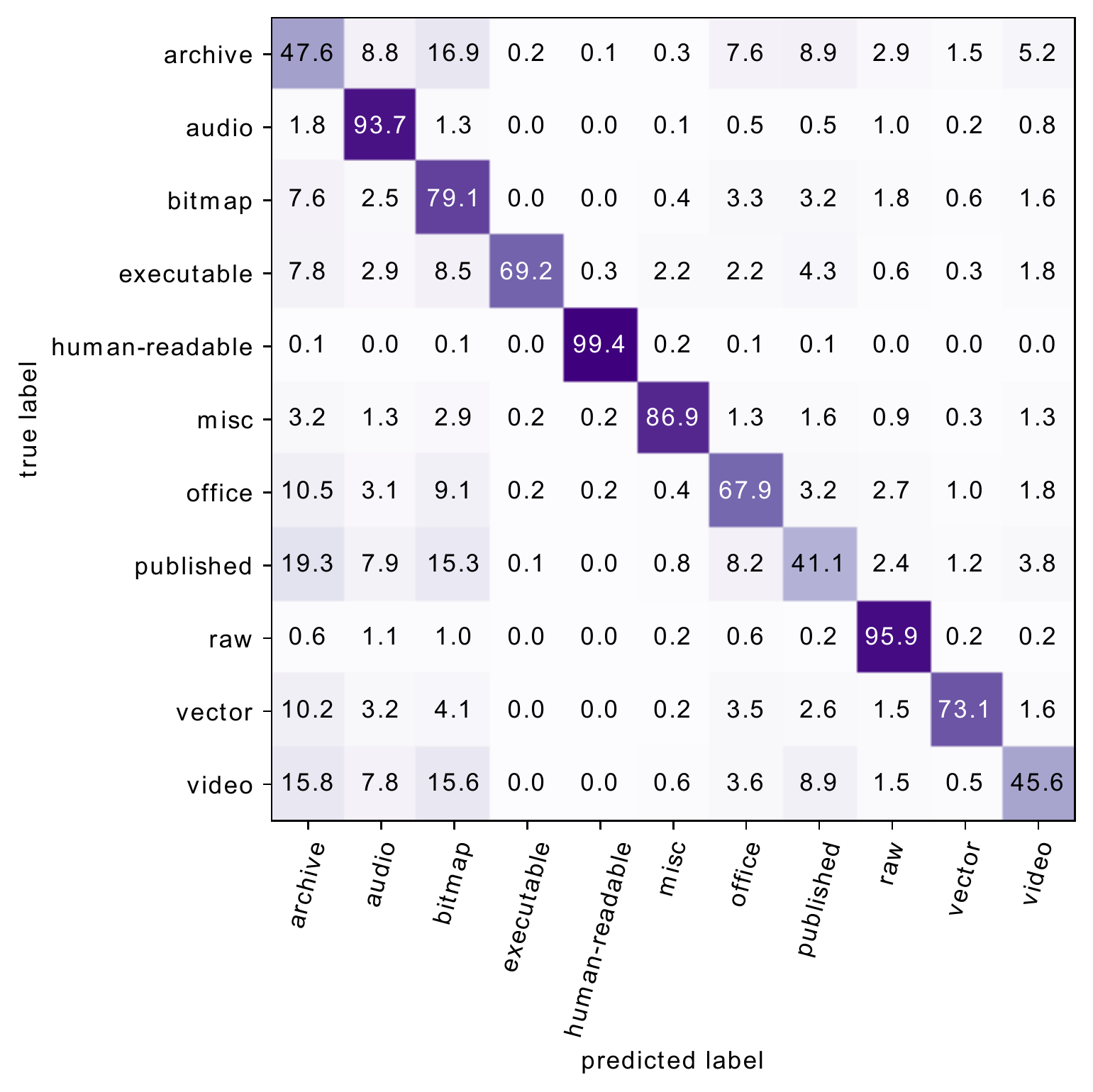}
                \caption{FiFTy (65.66\%)}
                \label{fig:fifty_512}
        \end{subfigure}%
        \begin{subfigure}[h!]{0.30\textwidth}
                \includegraphics[width=\linewidth]{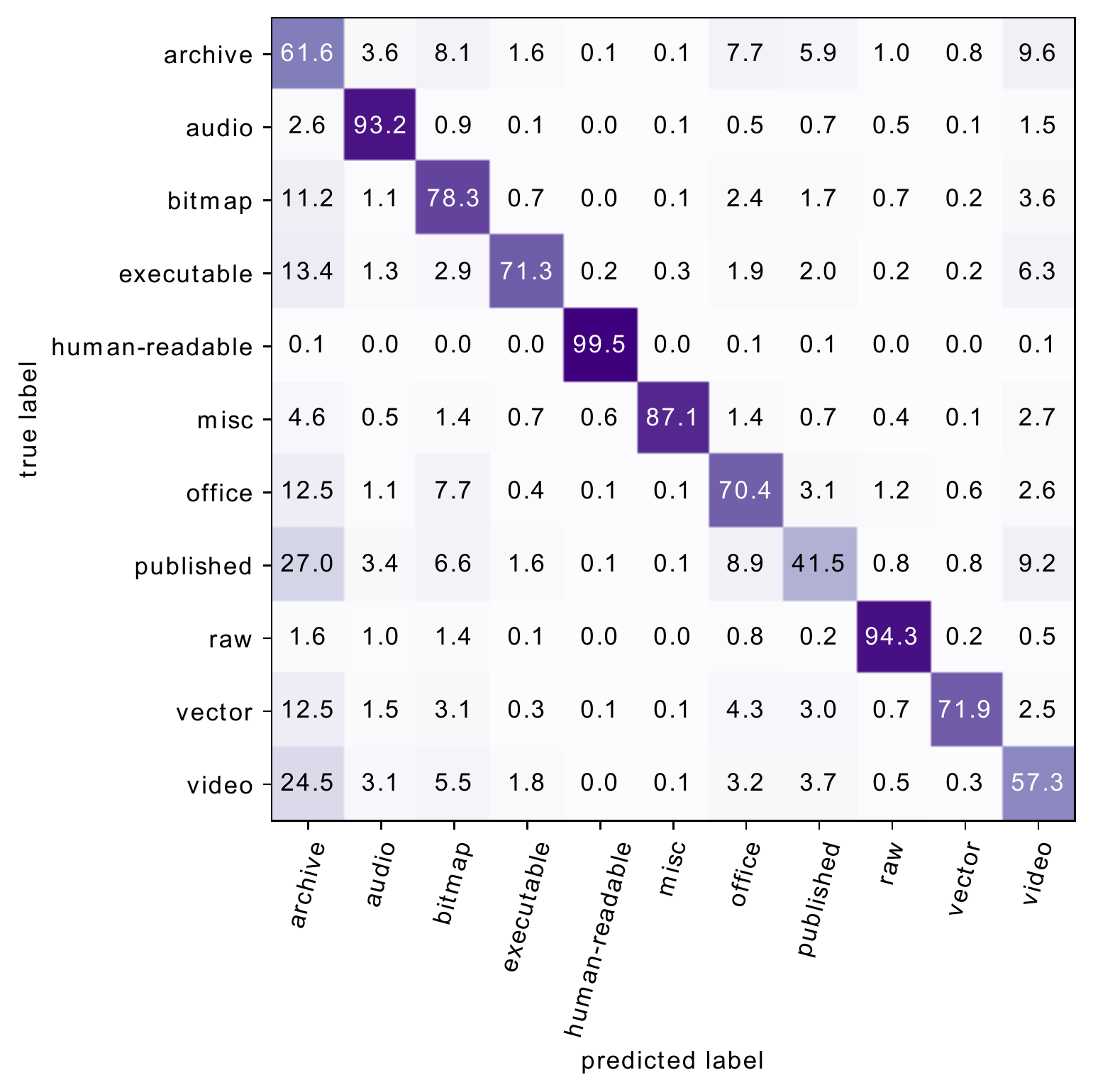}
                \caption{Baseline (67.5\%)}
                \label{fig:baseline_512}
        \end{subfigure}%
        \caption{Confusion matrices for Our models, FiFTy\cite{mittal2020fifty} and Baseline RNN model on 4 KB fragments in (a)-(e) and 512 bytes in (f)-(j). Due to large number of classes(75), classes belonging to same super class were clustered into one.}
        \label{fig:confusion_matrix}
\end{figure*}

\textbf{Accuracy.}
The confusion matrix of file types grouped by use-cases for Scenario 1 is plotted in Figure \ref{fig:confusion_matrix} (The grouping is listed in Table \ref{tab:grouping}). Most of the misclassification occurred in archived file group due to other file types embedded in them. The accuracy of the proposed models and FiFTy is listed in Table \ref{tab:Comp_our_fifty_acc}. In general, it can be observed that the accuracy of our models is comparable to FiFTy. However, our models have faster inference time.

\begin{table}[]
   \caption{Comparison between FiFTy and our models in terms of accuracy}
    \centering
    \begin{adjustbox}{max width=0.47\textwidth}
    \begin{tabular}{|c c l l l l l |}
        \hline
         \textbf{Scenario} & \textbf{\#of} & \textbf{Fragment} &  \multicolumn{4}{c|}{\textbf{Accuracy}}   \\
        &\textbf{files} &\textbf{Size} & \textbf{DSC}& \textbf{DSC-SE}& \textbf{M-DSC} & \textbf{FiFTy} \\
        \hline
        \multirow{2}{1em}{1} & \multirow{2}{2em}{75} & 4096 & 78.45 &\textbf{79.27}&78.68 & 77.04 \\
        & & 512 & 65.89& \textbf{66.33} & 64.04 & 65.66 \\
        \hline
        \multirow{2}{1em}{2} & \multirow{2}{2em}{11} & 4096 & 85.7 &87.10 & 85.33 & \textbf{89.91} \\
        & & 512 & 75.84 & 74.99 &72.18& \textbf{78.97} \\
        \hline
        \multirow{2}{1em}{3} & \multirow{2}{2em}{25} & 4096 & 93.06 & 93.32 &91.87& \textbf{94.64} \\
        & & 512 & 80.79 & 80.79&78.53&\textbf{87.97} \\
        \hline
        \multirow{2}{1em}{4} & \multirow{2}{2em}{5} & 4096 & 94.17&\textbf{94.61}&92.66 & 94.03 \\
        & & 512 & 87.14 & 87.32 & 83.41 & \textbf{90.30} \\
        \hline 
        \multirow{2}{1em}{5} & \multirow{2}{2em}{2} & 4096 & 99.28& \textbf{99.37} &99.24 & 99.12 \\
        & & 512 & 98.94 & 98.96 & 98.94 & \textbf{99.07} \\
        \hline
        \multirow{2}{1em}{6} & \multirow{2}{2em}{2} & 4096 & 99.59 & \textbf{99.69} & 99.62 & 99.59 \\
        & & 512 & 98.76 & 98.65 & 98.87 & \textbf{99.23} \\
        \hline
    \end{tabular}
    \end{adjustbox}
    \label{tab:Comp_our_fifty_acc}
\end{table}

\subsection{Discussion}

In summary, the results show that all three models have similar inference times, with minimal variance between runs. This is in contrast to the results reported in FiFTy \cite{mittal2020fifty}, where the inference time of FiFTy can be significantly larger. Our models are optimized for efficient inference by utilizing cutting-edge techniques to minimize computational overhead and reduce latency. As a result, we were able to achieve inference times that are comparable to state-of-the-art models while still maintaining high accuracy. This demonstrates the effectiveness of our proposed models in real-world applications where fast and efficient inference is crucial.

Similar to previous work \cite{mittal2020fifty,chen_hui_2018}, we observe that files with high entropy are difficult to classify because there is no statistical trace that the convolutional kernel can extract. Moreover, many files are container types that contain other files as embedded objects, e.g., \textit{pdf} files that can contain embedded \textit{jpg} images. As a result, classifiers behave erratically. Finally, similar file types with different format, e.g., \textit{ppt}, \textit{pptx}, and \textit{key}, are misclassified amongst themselves. However, \textit{doc} and \textit{docx} are not affected by this as \textit{docx} uses \textit{XML} whereas \textit{doc} is stored as binary.


\section{Related Work}
Several techniques have been proposed for identifying and classifying file fragments. The techniques range from fundamental methods utilizing magic numbers and file headers \cite{libmagic} to more advanced methods using machine learning and deep learning \cite{mittal2020fifty,beebe2013sceadan}. Generally, file carving methods can be broadly grouped into three categories: 1) statistical methods, 2) machine learning methods, and 3) deep learning methods. 
\textbf{Statistical methods.}
    Karresand et al. proposed \textit{Oscar} for determining the probable file type of binary data fragment \cite{Karresand2006OscarF}. \textit{Oscar} creates vector models based on the Byte Frequency Distribution (BFD) of 4096-byte-sized fragments from different file types. A vector distance between the mean and standard deviation of the segment to be classified and the centroids is calculated. If the distance is lower than a threshold value, the segment is classified as a modeling file. \textit{Sceadan} is a prominent open-source tool for file carving that utilizes a wide range of statistical features including uni-grams, bi-grams, entropy, mean byte value, and longest streak, among others \cite{beebe2013sceadan}. Ten separate input vectors are generated from these statistical features, which are then divided into four sets: uni-grams, bi-grams, all global features aside from n-grams, and a subset of global features. These sets are provided as inputs to the classification models. Ahmed et al. \cite{ahmed_lhee_shin_hong_2011} proposed an approach to identify file type based on its content. To reduce the computation time for identification, two techniques are applied. First, a subset of features based on occurrence frequency is selected. Second, file blocks of a 100-byte size are sampled. It was observed that this classifier performed well n low entropy file fragments (such as plain-text files, uncompressed images, etc.) and failed on high entropy file fragments(such as compressed files, binary executables, etc.). 

    \textbf{Machine learning methods}
    In \cite{Q_Li_SVM}, Li et al. proposed a Support Vector Machine (SVM) model that is trained using a feature vector which is based on the histogram of the data bytes. The SVM is utilized for binary classification, where one class is fixed as JPEG and the other class varies, including DLL, EXE, PDF, and MP3. Similarly, Fitzgerald et al. \cite{fitzgerald_mathews_morris_zhulyn_2012} presented an SVM model for fragment classification, where file fragments were treated as a bag of bytes represented by a feature vector that included uni-gram and bi-gram counts, as well as statistical measures such as entropy. Zheng et al. \cite{Zheng2015AFC} proposed an SVM-based approach that used BFD or histogram and entropy as feature vectors.
    Amirani et al. \cite{Amirani2013FeaturebasedTI} proposed a hierarchical combination of Principal Component Analysis (PCA) and Multi-layer Perceptron (MLP) for feature extraction. The extracted features were fed to the classifier. PCA selected N\textsubscript{1} number of features from BFD of the raw features. These N\textsubscript{1} features were used to train an auto-associative neural network that extracts N\textsubscript{2} features (N\textsubscript{2}$<$N\textsubscript{1}).
Bhatt et al. \cite{bhatt2020hierarchy} proposed a hierarchical classification approach for file fragment classification utilizing SVM as base classifiers. It was observed that the features used in the study are not adequate for complex file types, as they do not characterize the membership relationship between the file fragment and its parent class. The authors suggest refining the hierarchy branch for complex files and engineering features that better capture the membership relationship of a file fragment with its parent file type.
Haque and Tozal \cite{haque2022byte} proposed Byte2Vec, a novel feature generation model that extends the word2vec concept to map file fragments to dense vector representations. Byte2Vec generates vector representations utilizing the Skip-gram model, and a k-Nearest Neighbors (kNN) classifier is trained on these representations to identify fragment types. Byte2Vec corpus model works for various block sizes and file types. Using a publicly available dataset, the authors combine Byte2Vec with the kNN algorithm (Byte2Vec+kNN) for feature extraction and classification.

\textbf{Deep learning methods} Wang et al. \cite{wang_su_song_2018} proposed a one layer convolution with multiple kernel sizes. The raw byte values were converted to their corresponding binary representation and fed to the embedding layer to convert the binary value to a continuous dense vector. The authors studied 20 file types from the \textit{GovDocs1} dataset. Chen et al. \cite{chen_hui_2018} took a different approach and converted the 4096-byte file fragment to a 64x63 gray-scale image. The intuition was that data fragments from different files would have different texture features, reflected in the gray-scale image. The gray-scale images were fed to a deep CNN network like VGG \cite{DBLP:journals/corr/SimonyanZ14a} with many convolutions and max-pool layers followed by dense classification. Hiester \cite{Hiester2018FileFC} compared feed-forward, recurrent and convolutional neural networks with input fragments in the form of binary representation (fragment bits). The investigated file types were JPEG, GIF, XML and CSV. On these easily distinguishable file types, only recurrent networks gave satisfactory results. They emphasized lossless representation for achieving high accuracy, but this binary representation increases the input's dimensionality. 

    Mittal et al. \cite{mittal2020fifty} proposed \textit{FiFTy}, an open-source convolutional neural network for file fragment type identification. An open-source dataset was also developed by them, which is reported to be the largest open source dataset for file fragment classification. A compact neural network was developed that used trainable embedding space and convolutional neural networks. 
    When compared with neural network-based classifier \cite{mittal2020fifty,wang_su_song_2018,chen_hui_2018} our model achieves better accuracy on largest number of file types. 
    We have only compared our models with deep learning based methods as they can be accelerated using GPUs. Our models far exceed recurrent neural network based classifier \cite{Hiester2018FileFC} by being 660x faster. Compared to machine learning based classifier \cite{beebe2013sceadan}, which takes 9 min/GB,  our model takes 0.05 min/GB.  
\section{Conclusion}
In this paper, we proposed light-weight file fragment classification models based on depthwise separable CNNs. The objective is to develop a  model with better inference time as compared to the best-of-the-art model, without compromising on accuracy. In particular, we were able to design classification models with around 100K parameters. The models were evaluated using FFT-75 dataset that includes different scenarios. The evaluation results show the proposed models perform faster than the state-of-the-art file fragment classification model for all scenarios without dramatically degrading the accuracy. Several improvements can be made to increase the classification accuracy for classes with high misclassification rate. Without hardware constraints, neural architecture search \cite{zoph2018learning,pham2018efficient} is the best method to design an architecture for data-specific models. Moreover, for simple scenarios like classifying JPEGs against other file types, redundant connections in the neural network can be removed using in-network distillation \cite{hinton2015distilling,tian2019contrastive}.
\section{Acknowledgement}
The authors would like to acknowledge the Interdisciplinary Research Center for Intelligent Secure Systems at KFUPM. 

\bibliographystyle{splncs04}
\bibliography{main}



\end{document}